\documentclass[twocolumn, english, prd,superscriptaddress,
aps, showpacs, showkeys, amsmath, amssymb, floatfix, nofootinbib]{revtex4-1}

\pdfoutput=1
\usepackage[pdftex]{graphicx}
\usepackage[pdftex]{color}
\usepackage[colorlinks,bookmarks]{hyperref}
\usepackage{multirow}
\usepackage[export]{adjustbox}
\definecolor{linkblue}{rgb}{0,0,0.8}
\definecolor{linkgreen}{rgb}{0,0.5,0}
\hypersetup{linkcolor=linkblue, citecolor=linkgreen, urlcolor=linkblue}

\usepackage[T1]{fontenc}
\usepackage[utf8]{inputenc}
\usepackage{lmodern}
\usepackage{babel}
\usepackage{bm}
\usepackage{esint}
\usepackage{verbatim}
\usepackage[us]{datetime}
\usepackage{booktabs}
\usepackage{enumerate}
\usepackage{graphics}
\usepackage{float}
\usepackage{amsmath}
\usepackage{amsfonts}
\usepackage{amssymb}
\usepackage[normalem]{ulem}
\graphicspath{{./}}

\definecolor{vale}{rgb}{0,0.5, 1.}

\begin{document}
\title{Null test for interactions in the dark sector}

\author{Rodrigo von Marttens}
\affiliation{D\'epartment de Physique Th\'eorique and Center for Astroparticle Physics, Universit\'e de Gen\`eve, Quai E. Ansermet 24, CH-1211 Gen\`eve 4, Switzerland}
\affiliation{Departamento de Astronomia, Observat\'orio Nacional, 20921-400, Rio de Janeiro, RJ, Brasil}

\author{Valerio Marra}
\affiliation{Núcleo Cosmo-ufes \& Departamento de Física, Universidade Federal do Espírito Santo, 29075-910, Vitória, ES, Brasil}

\author{Luciano Casarini}
\affiliation{International Institute of Physics, Universidade Federal do Rio Grande do Norte, 59078-970 Natal, RN, Brasil}
\affiliation{Institute of Theoretical Astrophysics, University of Oslo, 0315 Oslo, Norway}

\author{J. E. Gonzalez}
\affiliation{Departamento de Astronomia, Observat\'orio Nacional, 20921-400, Rio de Janeiro, RJ, Brasil}
\affiliation{Departamento de F\'{\i}sica, Universidade Federal de Sergipe, 49100-000, Aracaju, SE, Brasil}

\author{Jailson Alcaniz}
\affiliation{Departamento de Astronomia, Observat\'orio Nacional, 20921-400, Rio de Janeiro, RJ, Brasil}
\affiliation{Departamento de F\'isica, Universidade Federal do Rio Grande do Norte, 59072-970, Natal, RN, Brasil}

\begin{abstract}

Since there is no known symmetry in Nature that prevents a non-minimal coupling between the dark energy (DE) and cold dark matter (CDM) components, such a possibility constitutes an alternative to  standard cosmology, with its theoretical and observational consequences being of great interest. In this paper we propose a new null test on the standard evolution of the dark sector based on the time dependence of the ratio between the CDM and DE energy densities which, in the standard $\Lambda$CDM scenario, scales necessarily as $a^{-3}$. We use the latest measurements of type Ia supernovae, cosmic chronometers and angular baryonic  acoustic oscillations to reconstruct the expansion history using model-independent Machine Learning techniques, namely, the Linear Model formalism and Gaussian Processes. We find that while the standard evolution is consistent with the data at $3\sigma$ level, some deviations from the $\Lambda$CDM model are found at low redshifts, which may be associated with the current tension between local and global determinations of $H_0$. 

%in the dark sector and, therefore, such possibility as well as its cosmological consequences must be explored. Testing a possible non-minimal coupling between the dark  matter  and  dark  energy  fields  constitutes  an  important  task  for  cosmology  and  fundamental  physics. Unless some unknown symmetry in nature prevents or suppresses a nonminimal coupling in the dark sector, the dark energy field may interact with the pressureless component of dark matter. Relaxing the conventional assumption of a minimal coupling between the dark energy and dark matter field introduces significant changes in the predicted universe's evolution.

\end{abstract}


\date{{\today}}
\maketitle

%%%%%%%%%%%%%%%%%%%%%%%%%
%%%%%%%%%%%%%%%%%%%%%%%%%
\section{Introduction}

According to the standard model of cosmology about 5\% of the energy content of the universe is made of particles belonging to the standard model of particle physics. The remaining 95\% is attributed to the so-called dark sector.
Roughly  25\% are thought to consist of a yet-undetected cold dark matter component (CDM), while dark energy (DE), the fuel that drives the current cosmic acceleration, would be responsible for the missing 70\%.
The fact that DE and CDM have comparable energy densities today -- the so-called coincidence problem -- has motivated the study,  at great depth, of dynamical models of DE that feature interactions between dark energy and dark matter~\cite[see e.g.][and references therein]{Copeland:2006wr,2010deto.book.....A,Li:2011sd,Chen:2010dk}, hoping to shed light on the nature of the dark sector. In this context, an important topic of research that has been extensively explored relies on considering some specific models for this interaction between the dark components in order to assess its cosmological consequences~\cite{Amendola:1999er,CalderaCabral:2008bx,Alcaniz:2012mh,Marttens:2016cba,Marra:2015iwa,Zimdahl:2014jsa,Gonzalez:2018rop}. 

Here, we adopt a different approach, i.e., instead of constraining the interaction parameter of a specific model we present a model-independent way to investigate whether or not such a interaction in the dark sector really exists.
For this, we introduce a new null test that is sensitive to the existence of a possible interaction between the dark components. Equivalently, if this null test is failed, then one may suspect that there may be new physics beyond the standard model and, in particular, that  dark matter and dark energy are not independent entities.
In other words, this null test has the ability to extract information that one may miss when the analysis performed is restricted to parameter estimation within a specific class of interacting dark energy models. 

The proposed null test is based on the time dependence of the ratio between CDM and DE energy densities, i.e., $r(z) = \rho_{\rm{CDM}}/\rho_{\rm{DE}}$, which in the $\Lambda$CDM model is given by $r\left(z\right)=r_{0}\left(1+z\right)^{3}$, where $r_{0}$ is the current value of this ratio. Since an interaction in the dark sector affects the dynamics of the components involved, this quantity is directly sensitive to the existence of such interaction. In order to carry out this new null test we will reconstruct the expansion history of the universe, in a model-independent way, using Machine Learning (ML) techniques applied to cosmic chronometers (CC) measurements, type Ia supernovae (SNe Ia) data and also angular Baryon Acoustic Oscillation (BAO) determinations.
In particular we will use the Linear Model formalism (LM) and Gaussian Processes (GP).
Regarding LM, we improve the so-called ``learning curve'' methodology by generalizing the ``Mean Square Error'' (MSE) and 
``Mean Square Prediction Error'' (MSPE) to the case of data that have an arbitrary covariance matrix and by taking into account the covariance matrix on the model parameters obtained from the training set. We call this generalization the ``calibrated learning curves''.

The outline of this paper is as follows. In Section~\ref{sec.int} we present the theoretical description of a rather general class of unified/interacting models, which is used to derive the $r\left(z\right)$ null test in Section~\ref{sec.null}. The Section~\ref{sec.data} is devoted to present the datasets used to perform the null test and to discuss about the priors on the further ``external'' parameters. In the Sections~\ref{sec:lin} and~\ref{sec.gp} the two methods used to perform the null test, i.e., LM and GP are discussed. The results are presented in Section~\ref{sec.res}, while Section~\ref{sec.conc} is dedicated to conclusions. The paper has also two appendices:  Appendix~\ref{conta} where we present the theoretical basis of the calibrated learning curves, and  Appendix~\ref{sec.lc} where the python script \texttt{learning\_curve} is released as an automatic tool to compute and plot the (calibrated) learning curves for any given dataset with a covariance matrix.\footnote{The python scripts can be downloaded from \href{https://github.com/rodrigovonmarttens/learning\_curve}{github.com/rodrigovonmarttens/learning\_curve}.}

%%%%%%%%%%%%%%%%%%%%%%%%%
%%%%%%%%%%%%%%%%%%%%%%%%%
\section{Interacting/Unified description}
\label{sec.int}

A simple and viable alternative to the standard cosmological model is to consider an interaction between the dark components of the universe.
The unknown nature of the dark sector does not allow us to provide a microphysical description 
of this interaction, which can only be modeled phenomenologically via a source term in the energy 
conservation equation,
\begin{align}
\dot{\rho}_{c}+3H\rho_{c}&=-Q \,, \label{cdm} \\
\dot{\rho}_{x}&=Q\,. \label{de}
\end{align}
From now on the subscripts $c$ and $x$ denote the dark matter and dark energy components, respectively, the dot denotes the derivative with respect to the cosmic time, and it was assumed that the dark energy equation of state (EoS) parameter is $w_{x}=-1$. %This is justified by the fact that the most recent observations constrain the DE EoS to be very close to $-1$ \cite{Abbott:2017wau,Aghanim:2018eyx}.
In the individual conservation equations above the interaction has opposite sign so that the total energy--momentum tensor is conserved.

Let us now assume that the interaction source can be parametrized via
\begin{equation}
Q=3H\gamma  R(\rho_{c},\rho_{x}) \,,
\end{equation}
where the dimensionless parameter $\gamma$ gives the interaction strength and the function $R$ specifies the type of interaction (see \cite{vonMarttens:2018iav} for details). It is then convenient to introduce the ratio $r$:
\begin{equation}
r\equiv\dfrac{\rho_{c}}{\rho_x}\quad\Rightarrow\quad \dot{r}=r\left(\dfrac{\dot{\rho}_{c}}{\rho_{c}}-\dfrac{\dot{\rho}_{x}}{\rho_{x}}\right)\,.
\label{r}
\end{equation}
Note that this quantity can be directly associated to the cosmic coincidence problem \cite{Velten:2014nra}.
Substituting equations (\ref{cdm}) and (\ref{de}) in the equation 
above, it is possible to obtain a differential equation for~$r$,
\begin{equation}
\dot{r}+3Hr\left(\gamma R \, \dfrac{\rho_{c}+\rho_{x}}{\rho_{c}\ \rho_{x}}+1\right)=0\,.
\label{rdot1}
\end{equation}

We are interested in the case in which the ratio between CDM and DE energy 
densities depends only on the scale factor (or, equivalently, on the redshift) and we assume, as an \textit{ansatz}, that the first term in the parenthesis is a function of $r$, that is,
\begin{equation}
\dot{r}+3Hr\big[ \gamma f\left(r\right)+1\big]=0\,.
\label{rdot2}
\end{equation}
This formalism to describe interactions in the dark sector 
was introduced in \cite{vonMarttens:2018iav} and is at the core of the null test proposed in this work.

If equation (\ref{rdot2}) has a solution that depends 
only on the scale factor, then we have an interacting model that can be 
associated to a unified model, i.e., we can combine the CDM and DE 
components in order to describe a single dark fluid. For this dark 
fluid, we define its energy density and pressure as the sum of the 
energy densities and pressures of CDM and DE,
\begin{align}
\rho_{d}&\equiv \rho_{c}+\rho_{x} \,, \label{rho1}\\
p_{d}&\equiv p_{c}+p_{x}=p_{x}\,. \label{p1}
\end{align}
In equation \eqref{rho1} one can express the unified dark energy density in terms of $r$ and only one of the energy densities of the dark sector's components:
\begin{equation}
\rho_{d}=\rho_{c}\left(1+\dfrac{1}{r}\right)\qquad\mbox{or}\qquad \rho_{d}=\rho_{x}\left(1+r\right)\,,
\label{rhod}
\end{equation}
and equation \eqref{p1}, using the second 
equation in (\ref{rhod}), can be rewritten as:
\begin{equation}
p_{d}= w_d \, \rho_{d} 
\qquad \text{with} \qquad w_d \equiv - \dfrac{1}{1+r}
\,,
\label{eosdark}
\end{equation}
where we defined the dark EoS parameter $w_d$.
Note that equation (\ref{eosdark}) is completely general, but it describes a unified model only 
if $r=r\left(a\right)$.
Since this dark fluid must be conservative,  energy conservation 
must be satisfied,
\begin{equation}
\dot{\rho}_{d}+3H\left(1+ w_d\right)=0\,.
\label{dotrhod}
\end{equation}

This unified description for the dark sector has been extensivly explored in the literature \cite{Zimdahl:2010tt,Zimdahl:2011mb,Marttens:2017njo}.

%%%%%%%%%%%%%%%%%%%%%%%%%
%%%%%%%%%%%%%%%%%%%%%%%%%
\section{The null test $r_{0}(z)$}
\label{sec.null}

For convenience, from now on, we will use the redshift $z$ instead the scale factor $a$.
The $\Lambda$CDM model is recovered if
\begin{equation}
r (z )=r_{0} (1+z )^{3} \qquad \text{with} \qquad r_{0}= \frac{\Omega_{c0}}{\Omega_{x0}} \,, \label{standard}
\end{equation}
and, in order to obtain a null test for interactions in the dark sector, we use \eqref{eosdark} together with \eqref{standard} so that the Friedmann equation becomes:
\begin{equation}
\dfrac{H^{2}}{H_{0}^{2}}=\Omega_{d0}\dfrac{1 + r_{0}\left(1+z\right)^{3}}{1+r_{0}}+\Omega_{b0}\left(1+z\right)^{3}\,,
\label{friedmann}
\end{equation}
where we have assumed spatial flatness, so that $\Omega_{d0}\equiv \Omega_{c0}+\Omega_{x0}=1-\Omega_{b0}$, and we have neglected radiation because we will consider only low-redshift data.
The above equation is the Friedmann equation for $\Lambda$CDM in 
terms of~$r_{0}$ and $\Omega_{d0}$.
Note that, since we are interested in an interacting scenario in which the 
interaction affects only the dark components, it is 
necessary to describe baryons and dark matter separately \cite{Marttens:2014yja}. 

The null test for interacting models is obtained solving equation \eqref{friedmann} for $r_{0}$:
\begin{equation}
r_{0}(z)=\dfrac{1-\Omega_{b0}+\Omega_{b0}\left(1+z\right)^{3}-H^{2}/H_{0}^{2}}{H^{2}/H_{0}^{2}-\left(1+z\right)^{3}}\,.
\label{r0z}
\end{equation}
Within the standard flat $\Lambda$CDM model one expects a constant $r_{0}(z) ={\Omega_{c0}}/{\Omega_{x0}} $. If a deviation is detected one may suspect not only that $\Lambda$CDM is falsified but also that dark matter and dark energy are not independent entities.

%%%%%%%%%%%%%%%%%%%%%%%%%
%%%%%%%%%%%%%%%%%%%%%%%%%
\section{Cosmological data}
\label{sec.data}

In order to carry out the null test of equation~\eqref{r0z}, it is necessary to determine the three quantities $H(z)$, $H_0$ and $\Omega_{b0}$.
The expansion history $H(z)$ will be reconstructed using  CC, SNe Ia and BAO data. The parameters $H_0$ and $\Omega_{b0}$ will be discussed in Section~\ref{subsec.ext}.

%%%%%%%%%%%%%%%%%%%%%%%%%
%%%%%%%%%%%%%%%%%%%%%%%%%
\subsection{Cosmic Chronometers}

\begin{figure}[t]
\centering
\includegraphics[width=\columnwidth]{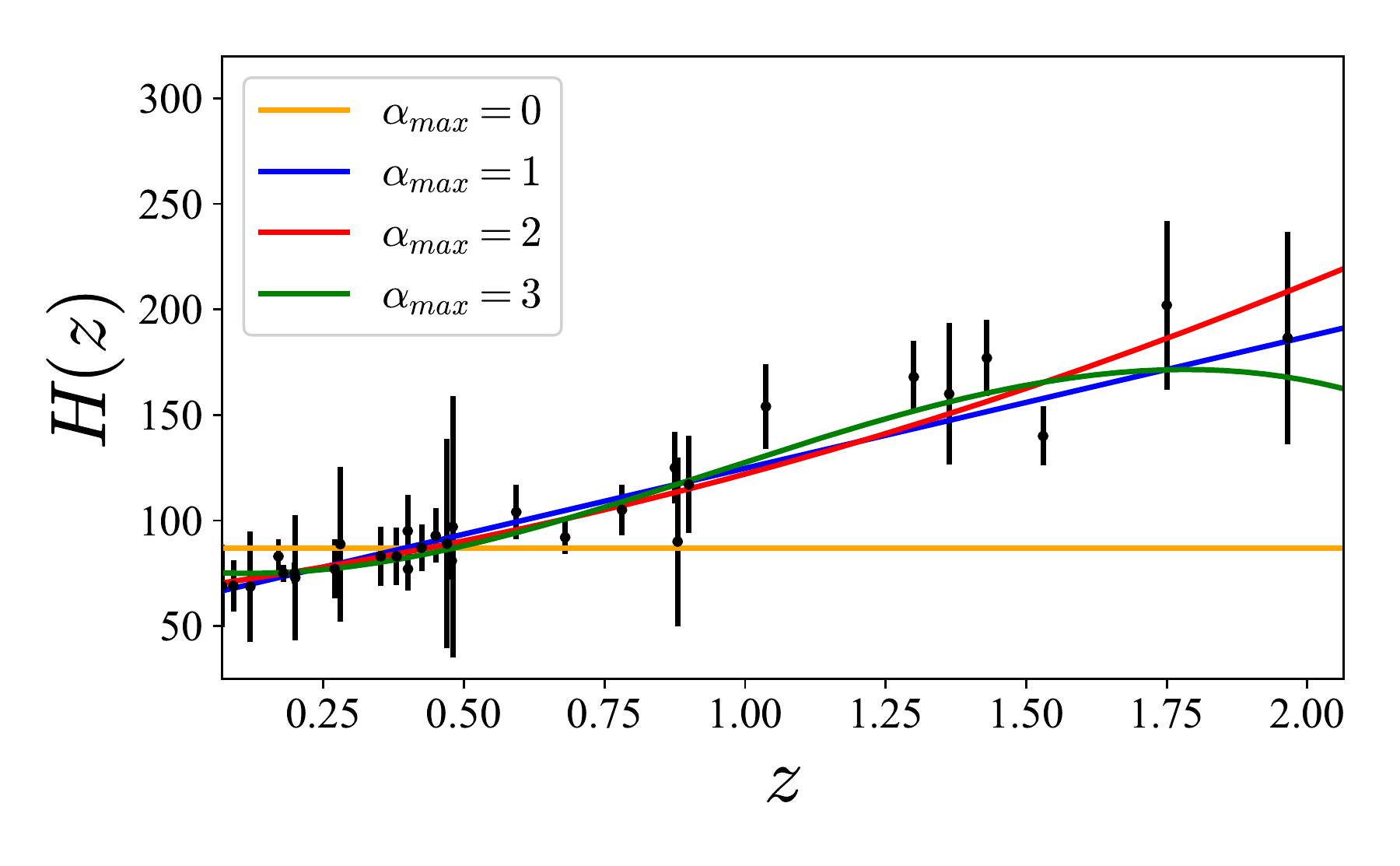} 
\caption{The black 
dots are the cosmic chronometer data. The solid lines are the 
reconstructions of the Hubble function using the linear model formalism. The data is presented in~\cite[][Table I]{Marra:2017pst}.}
\label{fighz}
\end{figure}

Cosmic chronometers are passively evolving old galaxies whose redshifts are known, and the expansion history of the  universe can be inferred directly from their differential ages \cite{Zhang:2012mp,Simon:2004tf,Moresco:2012jh,Moresco:2016mzx,
Ratsimbazafy:2017vga,Stern:2009ep,Moresco:2015cya}.
Here we will adopt the latest data as presented in \cite[][Table I]{Marra:2017pst}.
Figure \ref{fighz} illustrates the data points as well as the fitted Hubble function 
obtained using the linear model formalism presented in Section~\ref{sec:lin} at the first, second and third order.

%%%%%%%%%%%%%%%%%%%%%%%%%
%%%%%%%%%%%%%%%%%%%%%%%%%
\subsection{Type Ia Supernovae}\label{TIaSn}

The second dataset that we use to reconstruct $H(z)$ is the compressed supernova Ia Pantheon compilation (40 bins) \cite{Scolnic:2017caz}.\footnote{All the data (binned and full), as well as their 
covariance matrices, can be downloaded from \href{https://github.com/dscolnic/Pantheon}{github.com/dscolnic/Pantheon}.}
Note that, since we are performing a null test, 
the fact of we are using the binned catalog is not a problem in the sense 
of favoring the $\Lambda$CDM model. 

Type Ia Supernovas provide determinations of the distance modulus $\mu$, whose theoretical prediction is related to the luminosity distance $d_{L}$ according to:
\begin{equation}
\mu\left(z\right)=5\log\left[\dfrac{d_{L}\left(z\right)}{1 \text{ Mpc}}\right]+25\,,
\label{mu}
\end{equation} 
where the luminosity distance is given in Mpc.
In the standard statistical analysis, one adds to the distance modulus the nuisance parameter $M$, an unknown offset sum of the supernova absolute magnitude (and other possible systematics), which is degenerate with $H_{0}$.
In this analysis, as will be discussed in more 
details in section \ref{subsec.ext}, the value of $M$ is related to the prior on $H_{0}$. As we are assuming spatial flatness, the luminosity distance is related to the comoving distance $\mathcal{D}$ via
\begin{equation}
d_{L}\left(z\right)=\dfrac{c}{H_{0}}\left(1+z\right)\mathcal{D}\left(z\right)\,,
\label{dl}
\end{equation}
where $c$ is the speed of light, so that, using \eqref{mu}, one obtains:
\begin{equation}
\mathcal{D}\left(z\right)=\dfrac{H_{0}}{c}\left(1+z\right)^{-1}10^{\frac{\mu\left(z\right)}{5}-5}\,.
\label{Dsn}
\end{equation}
Finally, the normalized Hubble function $E\left(z\right)\equiv H\left(z\right)/H_{0}$ can be obtained by taking the inverse of the derivative of $\mathcal{D}(z)$ with respect to the redshift (denoted with a prime):
\begin{equation}
\mathcal{D}\left(z\right)=\int_{0}^{z}\dfrac{d\tilde{z}}{E\left(\tilde{z}\right)}\qquad\Rightarrow\qquad E\left(z\right)=\dfrac{1}{\mathcal{D}^{\prime}\left(z\right)}\,.
\label{Dz}
\end{equation}

\begin{figure}[t]
\centering
\includegraphics[width=\columnwidth]{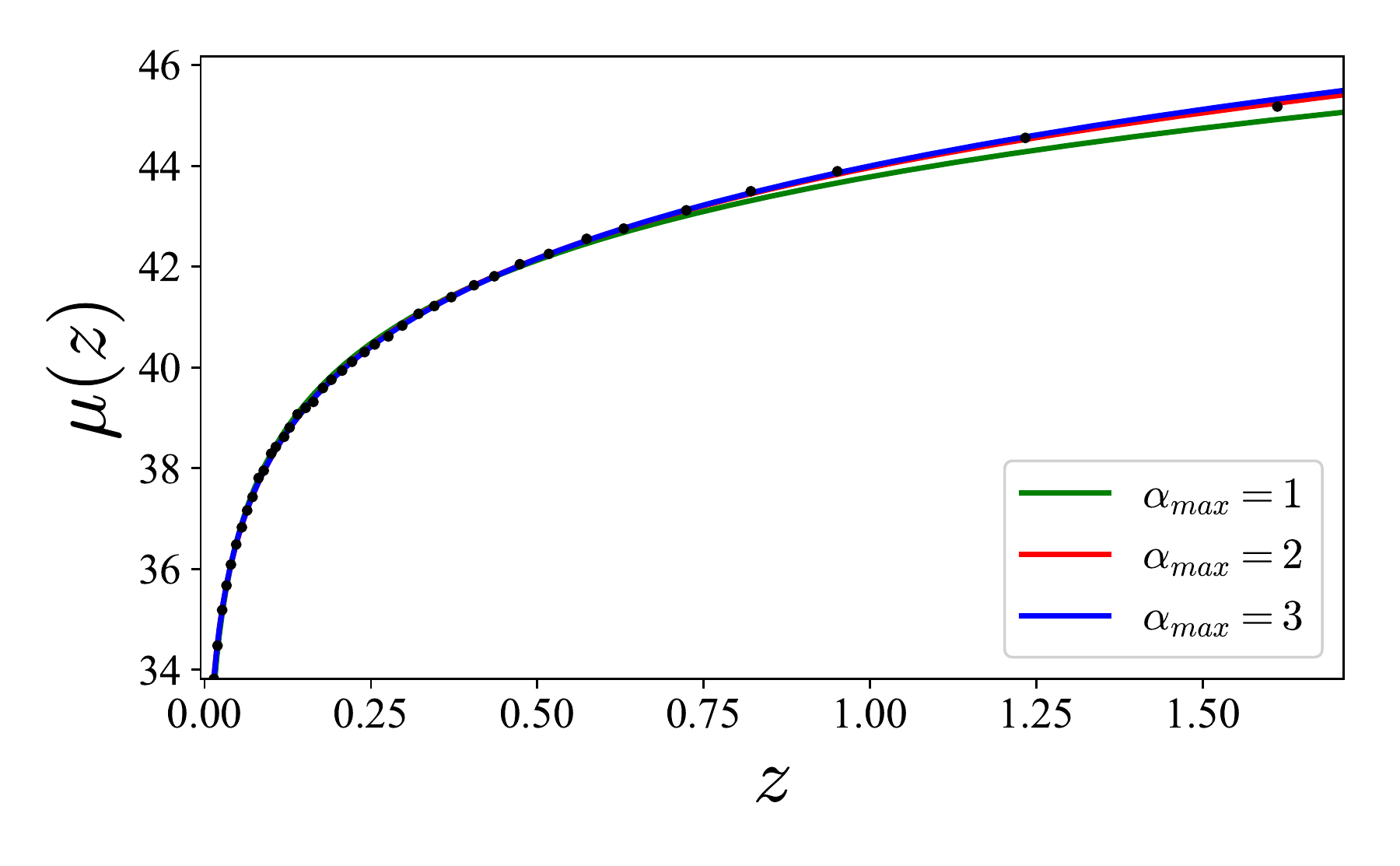} 
\caption{Distance modulus as a function of the redshift. The black 
dots are the Pantheon data (subtracting  $M=-19.25$). 
The solid lines are the 
reconstructions of the distance modulus using the linear model formalism.}
\label{figsn}
\end{figure}

The binned Pantheon data points (subtracting  $M=-19.25$) are 
shown in Figure \ref{figsn} together with the fitted distance modulus 
obtained using the linear model formalism at the first, second and third order. 

%%%%%%%%%%%%%%%%%%%%%%%%%
%%%%%%%%%%%%%%%%%%%%%%%%%
\subsection{BAO}

\begin{figure}[t]
\centering
\includegraphics[width=\columnwidth]{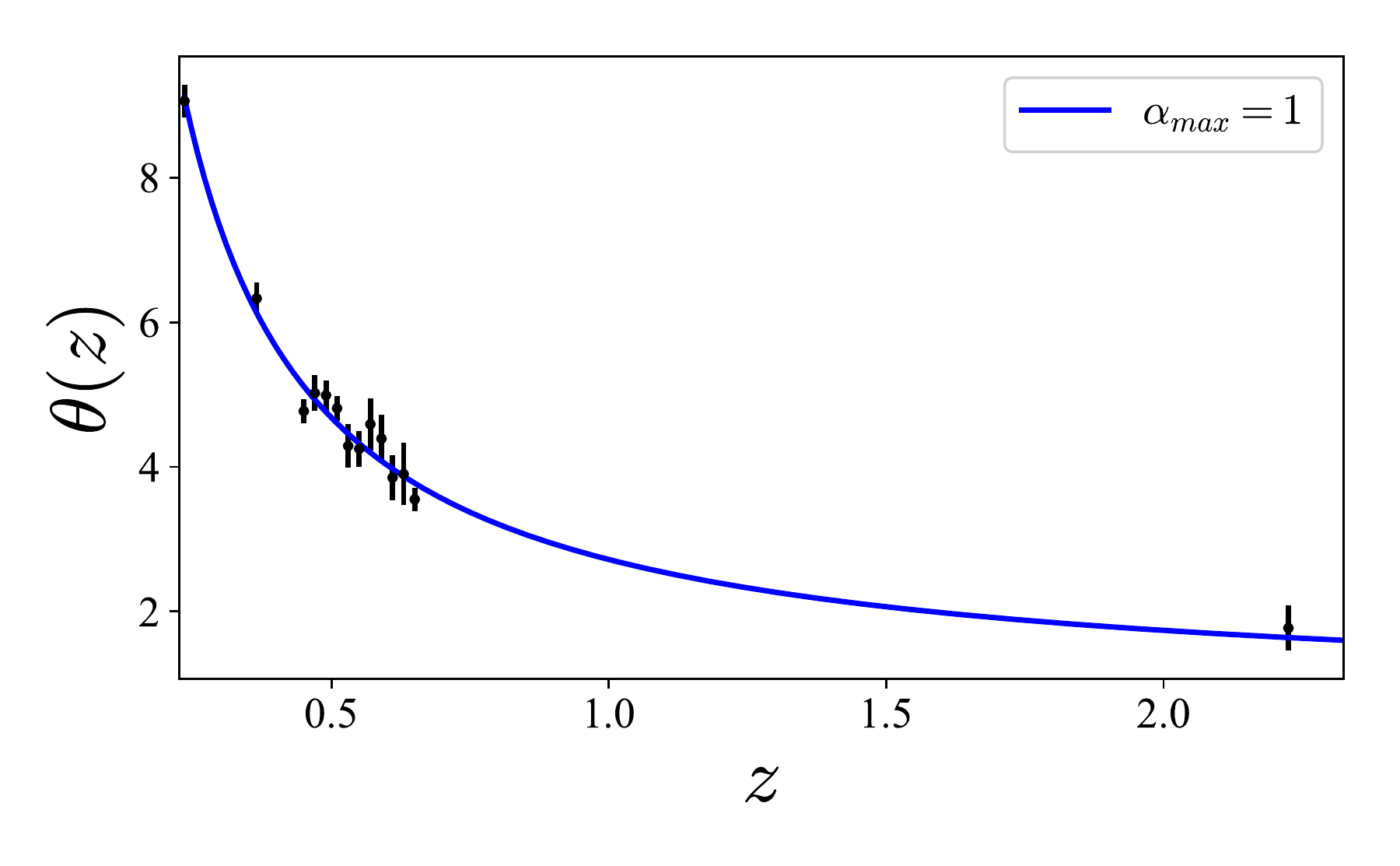} 
\caption{Angular BAO scale as a function of the redshift. The black 
dots are the data of Table~\ref{tabbao}. The solid line is the 
reconstruction of the BAO angular scale using the linear model formalism.}
\label{figbao}
\end{figure}

The last dataset that we use to reconstruct $H(z)$ are  model-independent angular  BAO 
determinations obtained using the angular correlation function \cite{Sanchez:2010zg}. In this case, we use 
14 uncorrelated data points  from \cite{Carvalho:2015ica,
Alcaniz:2016ryy,Carvalho:2017tuu,deCarvalho:2017xye}, which are presented in Table~\ref{tabbao} and illustrated in Figure~\ref{figbao} (with its first order fit). 
Model-independent determinations of the radial BAO scale were recently obtained in~\cite{Marra:2018zzo}.

\begin{table}[t]
\centering
\begin{tabular}{lllll}
\hline\hline
Catalog & $z$    & $\theta(z)$ & $\sigma_{\theta(z)}$ & Ref. \\
\hline
SDSS-DR7          & 0.235        & 9.06        & 0.23        & \cite{Alcaniz:2016ryy} \\
SDSS-DR7          & 0.365        & 6.33        & 0.22        & \cite{Alcaniz:2016ryy} \\
SDSS-DR10         & 0.450        & 4.77        & 0.17        & \cite{Carvalho:2015ica} \\
SDSS-DR10         & 0.470        & 5.02        & 0.25        & \cite{Carvalho:2015ica} \\
SDSS-DR10         & 0.490        & 4.99        & 0.21        & \cite{Carvalho:2015ica} \\
SDSS-DR10         & 0.510        & 4.81        & 0.17        & \cite{Carvalho:2015ica} \\
SDSS-DR10         & 0.530        & 4.29        & 0.30        & \cite{Carvalho:2015ica} \\
SDSS-DR10         & 0.550        & 4.25        & 0.25        & \cite{Carvalho:2015ica} \\
SDSS-DR11         & 0.570        & 4.59        & 0.36        & \cite{Carvalho:2017tuu} \\
SDSS-DR11         & 0.590        & 4.39        & 0.33        & \cite{Carvalho:2017tuu} \\
SDSS-DR11         & 0.610        & 3.85        & 0.31        & \cite{Carvalho:2017tuu} \\
SDSS-DR11         & 0.630        & 3.90        & 0.43        & \cite{Carvalho:2017tuu} \\
SDSS-DR11         & 0.650        & 3.55        & 0.16        & \cite{Carvalho:2017tuu} \\
SDSS-DR12Q$\quad$ & 2.225$\quad$ & 1.77$\quad$ & 0.31$\quad$ & \cite{deCarvalho:2017xye} \\
\hline\hline
\end{tabular}
\caption{Angular BAO data.}
\label{tabbao}
\end{table}

The theoretical BAO angular scale, in degrees, is given by,
\begin{equation}
\theta\left(z\right)=\dfrac{r_{s}}{d_{A}\left(z\right)\left(1+z\right)}\left(\dfrac{180}{\pi}\right)\,,
\label{theta}
\end{equation}
where $r_{s}$ is the sound horizon of the primordial photon-baryon fluid 
at the drag time and $d_{A}\left(z\right)$ is the angular diameter  distance, which, in a flat universe, is related to the comoving  distance by: 
\begin{equation}
d_{A}\left(z\right)=\dfrac{c}{H_{0}\left(1+z\right)}\mathcal{D}\left(z\right)\,.
\label{da}
\end{equation}
Substituting the equation above in equation (\ref{theta}), one obtains the 
following explicit relation between $\theta\left(z\right)$ and 
$\mathcal{D}\left(z\right)$,
\begin{equation}
\mathcal{D}\left(z\right)=\dfrac{H_{0}}{c}\dfrac{r_{s}}{\theta\left(z\right)}\left(\dfrac{180}{\pi}\right)\,.
\label{Dbao}
\end{equation}
The normalized Hubble function $E(z)$ is then obtained using equation~\eqref{Dz}.

%%%%%%%%%%%%%%%%%%%%%%%%%
%%%%%%%%%%%%%%%%%%%%%%%%%
\subsection{Further parameters \label{subsec.ext}} 

%As will be discussed further in more details, $H_{0}$ and $\Omega_{b0}$
%are external parameters that will be fixed, and their errors will be taken
%into account to compute the error of the reconstruction of $r_{0}$.

As mentioned earlier, in order to perform the null test of equation~\eqref{r0z}, it is necessary to determine also $H_0$ and $\Omega_{b0}$.%
\footnote{When using SNe Ia data or angular BAO measurements, only $\Omega_{b0}$ is necessary as one obtains directly $H/H_{0}$.}
We will adopt two different sets of values for these parameters. 
The first set is related to model-independent measurements: the local determination of 
$H_0$ obtained from low-redshift 
SN Ia data calibrated with loca Cepheids \cite{Riess:2018byc}, and 
the measurement of the baryon density parameter from big bang nucleosynthesis~\cite{Pettini:2012ph},
\begin{align}
H_{0}&=73.52\pm 1.62 \;  \frac{\text{km/s}}{\text{Mpc}} \,, \label{h0riess} \\
\omega_{b}&=0.0223\pm 0.0009 \,. \label{obriess}
\end{align}

The second set of values comes from the most recent results from Planck 
\cite{Aghanim:2018eyx}. In this work, we use the results obtained with 
TT,TE,EE+lowE+lensing+BAO,
\begin{align}
H_{0}&=67.66\pm 0.42 \;  \frac{\text{km/s}}{\text{Mpc}} \,, \label{h0planck}\\
\omega_{b}&=0.02242\pm 0.00014 \,. \label{obplanck}
\end{align}

\begin{figure}[t]
\centering
\includegraphics[width=\columnwidth]{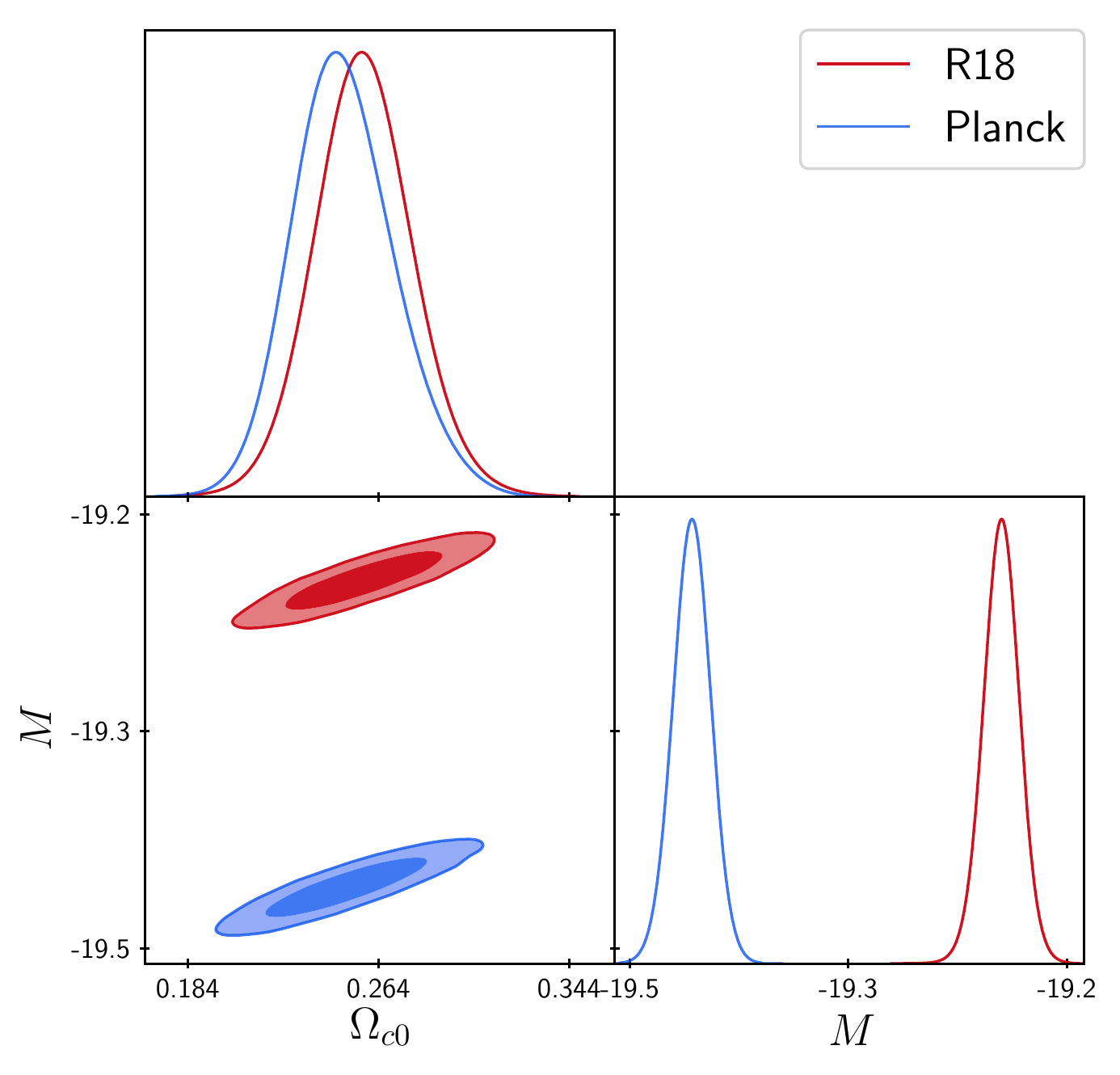} 
\caption{Statistical analysis for the $\Lambda$CDM model using the Pantheon sample of type Ia SN data. The result in red 
was obtained fixing $H_{0}$ to \eqref{h0riess} (model independent priors), and the blue result was obtained fixing $H_{0}$ to \eqref{h0planck}.}
\label{figM}
\end{figure}

When using SNe Ia, we have the additional nuisance parameter $M$.
Since $H_{0}$ is fixed according to (\ref{h0riess}) or 
(\ref{h0planck}), we choose the respective values of $M$ from a statistical analysis of the 
$\Lambda$CDM model with the Pantheon data obtained by fixing $H_{0}$ to the values 
previously mentioned. 
In order to perform the analysis, we have used the statistical code MontePython \cite{Audren:2012wb,Brinckmann:2018cvx}.
When $H_{0}$ is fixed to the local determination (\ref{h0riess}), we found the best-fit 
value  $M=-19.25$, while when $H_{0}$ is fixed to the Planck result (\ref{h0planck}) we obtained $M=-19.42$. The complete result of the statistical analysis with 
$1\sigma$  confidence levels is shown in Table~\ref{tabM} and illustrated in Figure~\ref{figM}.

\begin{table}[t]
\begin{tabular}{l|c|c|c|c} 
 \hline \hline
 & $H_{0}$ & $\omega_{b}$ & $M$ & $\Omega_{c0}$ \\ \hline 
R18/BBN$\quad$ & $73.52$ & $0.0223$ & $-19.25^{+0.01}_{-0.01}$ & $0.256^{+0.022}_{-0.023}$ \\ 
Planck & $67.66$ & $0.02242$ & $-19.42^{+0.01}_{-0.01}$ & $0.249^{+0.022}_{-0.023}$ \\ 
\hline\hline
 \end{tabular} \\ 
\caption{Result of the statistical analysis with the type Ia SN data from the Pantheon sample.}
\label{tabM}
\end{table}

When using the angular BAO determinations, one has the sound horizon $r_{s}$ as an additional parameter that 
must be specified. Maintaining the idea of the two sets of values we choose for the first set the 
model-independent result obtained from low-redshift standard rulers \cite{Verde:2016ccp} 
$r_{s}=\left(101.0\pm2.3\right)h^{-1}$ Mpc. Using the local measurement of $H_0$ of  (\ref{h0riess}), one has: 
\begin{equation}
r_{s}=146.6\pm4.1 \text{ Mpc} \,.
\end{equation}
For the second set of values, we use the same Planck determination from \cite{Aghanim:2018eyx}:
\begin{equation}
r_{s}=147.21\pm0.23 \text{ Mpc} \,.
\end{equation}
%

%%%%%%%%%%%%%%%%%%%%%%%%%
%%%%%%%%%%%%%%%%%%%%%%%%%
\section{Linear model formalism}
\label{sec:lin}

Here, we will describe how to reconstruct the cosmological functions, and also their derivatives, using the Linear Model formalism (LM); see, also, \cite{Marra:2017pst,Gomez-Valent:2018hwc,Pinho:2018unz}.

%%%%%%%%%%%%%%%%%%%%%%%%%%%%%%%%%%%%
\subsection{Linear Models} \label{lma}
%%%%%%%%%%%%%%%%%%%%%%%%%%%%%%%%%%%%

Let us choose a set of base functions $g_{\alpha}(z)$ whose {\it linear} combination will constitute the template function~$t(z,\{c_{\alpha}\})$:
\begin{equation} \label{tempe1}
t(z,\{c_{\alpha}\}) = \sum_{\alpha=0}^{\alpha_{\rm max}} c_\alpha \, g_{\alpha}(z) \,,
\end{equation}
where $\alpha$ is an integer.
The assumption is that $t(z,\{c_{\alpha}\})$ can describe the actual functions that we want to reconstruct: $H(z)$, $\mu(z)$ or $\theta(z)$. 
Clearly, this is conditional to an appropriate choice of $g_{\alpha}(z)$ and the order $\alpha_{\rm max}$. Usually, $g_{0}$ is a constant, often unity.
The template will have $\alpha_{\rm max}+1$ coefficients.

Let us then assume that the data are given by:
\begin{equation}
d_i = t_i + e_i \,,
\end{equation}
where $t_i=t(z_i, \{c_{\alpha}\})$ and $e_i$ are Gaussian errors with covariance matrix $\Sigma_{ij}$.

Next, we fit the template $t$ to the data and use the LM formalism to calculate the Fisher matrix relative to the parameters $c_\alpha$, which gives an exact description of the likelihood as the template is linear in its parameters.
The Fisher matrix is:
\begin{equation} \label{fisher}
F_{\alpha \beta} = g_{\beta i} \Sigma^{-1}_{i j} g_{\alpha j}
\end{equation}
where $g_{\alpha i}= g_{\alpha}(z_{i})$, and the best-fit values of $c_\alpha$ are:
\begin{equation}
c_{\alpha, \rm bf}= F^{-1}_{\alpha \beta} B_\beta \equiv \Sigma_{\alpha \beta} B_\beta  \,,
\end{equation}
where $B_\alpha = d_i \Sigma^{-1}_{i j} g_{\alpha j}$ and we defined the covariance matrix $\Sigma_{\alpha \beta}$ on the parameters.
Summarizing, this formalism had allowed us to exactly propagate the data covariance matrix $\Sigma_{ij}$ into the parameter covariance matrix~$\Sigma_{\alpha \beta}$.

%%%%%%%%%%%%%%%%%%%%%%%%%%%%%%%%%%%%
\subsection{Error on null test reconstruction} \label{error}
%%%%%%%%%%%%%%%%%%%%%%%%%%%%%%%%%%%%

The null test $r_0 (z , \{\theta_\alpha \})$ is a nonlinear function of the template parameters and of the additional parameters of Section~\ref{subsec.ext}.
The corresponding covariance matrix $S_{\alpha \beta}$ is obtained by forming an appropriate block diagonal matrix using the covariance matrices of the corresponding parameters (e.g., $\Sigma_{\alpha \beta}$).
As we have chosen independent data, correlations among different datasets are not expected to be important.

In order to compute the error on $r_0 (z , \{\theta_\alpha \})$ due to the uncertainty encoded in the covariance matrix $S_{\alpha \beta}$, a straightforward approach is to apply a change of variable from $\{\theta_{\alpha}\}$ to $r_0$.
At the first order, the error is then given by:
\begin{align} \label{linear}
\sigma^{2}_{r_0} = J_{\alpha}S_{\alpha \beta} J_{\beta} \,,
\end{align}
where
\begin{equation}
J_{\alpha} =\left. \frac{\partial r_0 (z , \{\theta_\alpha \})}{\partial \theta_{\alpha}} \right|_{\theta_{\alpha, \rm bf}} \,.
\label{eq:jacobian-matrix}
\end{equation}
%

%%%%%%%%%%%%%%%%%%%%%%%%%
%%%%%%%%%%%%%%%%%%%%%%%%%
\subsection{Base functions}\label{subsec.base}

In order to reconstruct $H(z)$, $\mu(z)$ and $\theta(z)$, we will adopt the following base functions, respectively:
\begin{align} 
t_{H}\left(z, \lbrace c_{\alpha}\rbrace\right) &= \sum_{\alpha} c_{\alpha}z^{\alpha}\,, \label{temph}\\
t_{\mu}\left(z, \lbrace c_{\alpha}\rbrace\right) &= \sum_{\alpha} c_{\alpha}\left[\ln z \right]^{\alpha}\,,\label{tempmu} \\
t_{\theta}\left(z, \lbrace c_{\alpha}\rbrace\right) &= \sum_{\alpha} c_{\alpha}z^{-\alpha}\,.  \label{temptheta}
\end{align}
In order to choose $\alpha_{\rm max}$ we will use the so-called ``learning curves'', a machine learning tool.

%%%%%%%%%%%%%%%%%%%%%%%%%
%%%%%%%%%%%%%%%%%%%%%%%%%
\subsection{Calibrated learning curves}

The availability of large datasets is increasingly a defining feature of modern cosmology, in which data analysis has become an important component. 
Computations that were not possible a few decades ago can now be performed on GPU-based laptops.
Machine learning includes a set of statistical techniques  that allows computer systems to learn from examples, data, and experience, rather  than following pre-programmed rules.

A simple method that is commonly used to choose the template order $\alpha_{\rm max}$ is the computation of the reduced chi-square~$\chi^2_{\nu}$:
\begin{equation} \label{soda}
\chi^2_{\nu} = \frac{\left(d_{i}-\bar{t}_i\right)\Sigma^{-1}_{ij}\left(d_{j}-\bar{t}_j\right)}{N_{\rm tot}-\alpha_{\rm max}-1}\,,
\end{equation}
where $d_{i}$ are the $N_{\rm tot}$ data of the full dataset $D$ with covariance matrix $\Sigma_{ij}$, and $\bar{t}_i = t(z_i, \{c_{\alpha, {\rm bf}}\})$ where $c_{\alpha, {\rm bf}}$ are the best-fit parameters.
If one finds that $\chi^2_{\nu}$ is compatible with its corresponding distribution with $N_{\rm tot}-\alpha_{\rm max}-1$ degrees of freedom, then the null hypothesis that $t(z,\{c_{\alpha}\})$ is the correct model is not rejected (it is ``ruled in''). While powerful in its simplicity, this method is somewhat subjective as it strongly depends on the $p$-value threshold (e.g., $p=0.01$) that one is supposed to use. For example, two or more values of $\alpha_{\rm max}$ could be acceptable.

In order to overcome this difficultly and extract more information from the data, we will study the learning curves.
These usually are used in contexts in which the data covariance matrix is not available and so a performance statistics with a known distribution (like $\chi^2_{\nu}$) cannot be built.
Therefore, we will have to first generalize the standard learning curves to what we call the ``calibrated learning curves.''

Let us then consider two disjoint subsets of the data set $D$ of $N_{\rm tot}$ elements: the training set~$d$ and the validation set~$\tilde{d}$.%
\footnote{From now on $d$ will refer to  the training set and not to the full data set.}
The basic idea behind the learning curves consists in using the training set to fit the model and then test the latter with the validation set.
Within machine learning the fit is usually obtained by minimizing the ``Mean Squared Error'' (MSE):
\begin{equation} \label{mse}
\overline{\text{MSE}} = \frac{1}{N} \sum_{i=1}^{N} \left(d_{i}- \bar{t}_i  \right)^2 \,,
\end{equation}
where $N$ is the number of data points $d_i$ and $\bar{t}_i = t(z_i, \{c_{\alpha, {\rm bf}}\})$, where $c_{\alpha, {\rm bf}}$ are the parameters that minimize the MSE. From now on we will denote the minimized $\overline{\text{MSE}}$ with just MSE.
The test on the validation set is then performed by computing the ``Mean Squared Prediction Error'' (MSPE):
\begin{equation} \label{mse}
\text{MSPE} = \frac{1}{\tilde N} \sum_{i=1}^{\tilde N} \left(\tilde d_{i}- \bar{t}_i  \right)^2 \,.
\end{equation}
The use of the new $\tilde{N}$ data points $\tilde d_{i}$ justifies the alternative name ``out-of-sample mean squared error''.
Note that $N_{\rm tot} \ge N + \tilde{N}$.

The learning curves are then the values of the MSE and the MSPE as a function of the  training-set size $N$ while keep the validation-set size $\tilde{N}$ fixed. Usually, $\tilde{N}$ is 20--30\% of $N_{\rm tot}$.
The expectation is that the MSE will increase as the same number of parameters will be fitted to more data, and the MSPE will decrease as the training will produce a more reliable fitted model. In particular:
\begin{itemize}

\item an under-fitting model will feature converging but high (poor) MSE and MSPE;

\item an over-fitting model will feature low MSE but high MSPE because the model is fitting the noise in the training set which is different with respect to the validation set;

\item an optimal model will feature converging and low MSE and MSPE. Moreover, the sooner the convergence is reached, the better. Indeed, if the MSE and MSPE converge at $N< N_{\rm tot}-\tilde{N}$ reaching a plateau, it means that there were enough data to optimally train the model.

\end{itemize}
For more details,  see, for instance  \cite{hastie_09_elements-of.statistical-learning,Abu-Mostafa,Goodfellow-et-al-2016,Gron:2017:HML:3153997}.

The MSE and the MSPE do not use the data covariance matrix and, therefore, it is difficult to assess statistically their values.
For example, it is not clear how to define ``high'', ``low'' and ``close to each other''.
Therefore, we will calibrate the learning curve method by introducing new performance estimators that can be interpreted quantitatively in a statistical way.

A natural alternative to the MSE is the reduced chi-square function $\chi^2_{\nu}$,
\begin{equation}
\chi^2_{\nu} = \frac{\left(d_{i}-\bar{t}_i\right)\Sigma^{-1}_{ij}\left(d_{j}-\bar{t}_j\right)}{N-\alpha_{\rm max}-1}\,,
\end{equation}
where $\Sigma$ is the covariance matrix of the training set $d$ and $\alpha_{\rm max}+1$ is the number of fitted parameters.
Assuming that  $t(z,\{c_{\alpha}\})$ is the correct model and that the data $d_{i}$ are distributed according to a multivariate 
Gaussian distribution of covariance matrix $\Sigma$, it is $\big\langle\chi^2_{\nu}\big\rangle=1$.

A natural alternative to MSPE is,
\begin{equation} \label{co1}
\tilde{\chi}^2_{\nu} =\frac{\left(\tilde{d}_{i}-\bar{t}_i\right)\tilde{\Sigma}^{-1}_{ij} \left(\tilde{d}_{j}-\bar{t}_j\right)}{\tilde{N}}\,,
\end{equation}
where $\tilde{\Sigma}$ is the covariance matrix related to the validation set $\tilde{d}$.
As these data were not used to obtain the best-fit parameters $\{c_{\alpha, {\rm bf}}\}$ the denominator only contains 
$\tilde{N}$.
However, the expected value of $\tilde{\chi}^2_{\nu}$ is not unity as $\bar{t}_{i}$ is not the true value. Consequently, $\chi^2_{\nu}$ and $\tilde{\chi}^2_{\nu}$ will not converge to the same numerical value.
Here, we propose a new generalization of the MSPE:
\begin{equation}
\tilde{\chi}^2_{\delta} = \frac{\left(\tilde{d}_{i}-\bar{t}_{i}\right)\tilde{\Sigma}^{-1}_{ij}\left(\tilde{d}_{j}-\bar{t}_{j}\right)}{\tilde{N}}-\frac{ \Sigma_{\alpha\beta}\tilde{\Sigma}^{-1}_{\alpha\beta}}{\tilde{N}} \,,
\end{equation}
whose expectation value is unity as discussed in Appendix~\ref{conta}.

In order to obtain smooth learning curves, we compute, for a fix $N$, $\chi^2_{\nu}$ and $\tilde{\chi}^2_{\delta}$ for 2000 partitions, from which we then compute mean and standard deviation.
Note that the performance estimators $\chi^2_{\nu}$ and $\tilde{\chi}^2_{\delta}$ have an expectation value of unity independently of the training set size $N$, but this is true only if the expectation value is taken using independent training sets while here the training sets all come from the \textit{same} dataset.
In other words, the 2000 partitions are used to extract the average behavior of a training set of size $N$ from the full dataset $D$.

For smaller $N$ it is quite likely to obtain low values of $\chi^2_{\nu}$ as its distribution is skewed towards lower values, while for larger $N$ one expects $\chi^2_{\nu}$ and $\tilde{\chi}^2_{\delta}$ to converge to the common value of unity.
If they converge reaching a plateau it means that $t(z,\{c_{\alpha}\})$ is the correct model and that the latter has been trained optimally by the data.

In our analysis, we adopt the following criterion in order to choose the best order $\alpha_{\rm max}$: the optimal $\alpha_{\rm max}$ is the one for which $\chi^2_{\nu}$ and $\tilde{\chi}^2_{\delta}$ converge fastest to unity with a plateau.
It is very important to emphasize that this learning curve procedure is completely 
independent of any physical assumption, depending only on data.

%%%%%%%%%%%%%%%%%%%%%%%%%
\subsection{Learning curve results}
%%%%%%%%%%%%%%%%%%%%%%%%%
In the following, we present the results of the learning curve analysis for the datasets of Section~\ref{sec.data}.

%%%%%%%%%%%%%%%%%%%%%%%%%
\subsubsection{Cosmic chronometers}
\begin{figure}[t]
\centering
\includegraphics[width=\columnwidth, trim ={0 2.44cm 0 0}, clip]{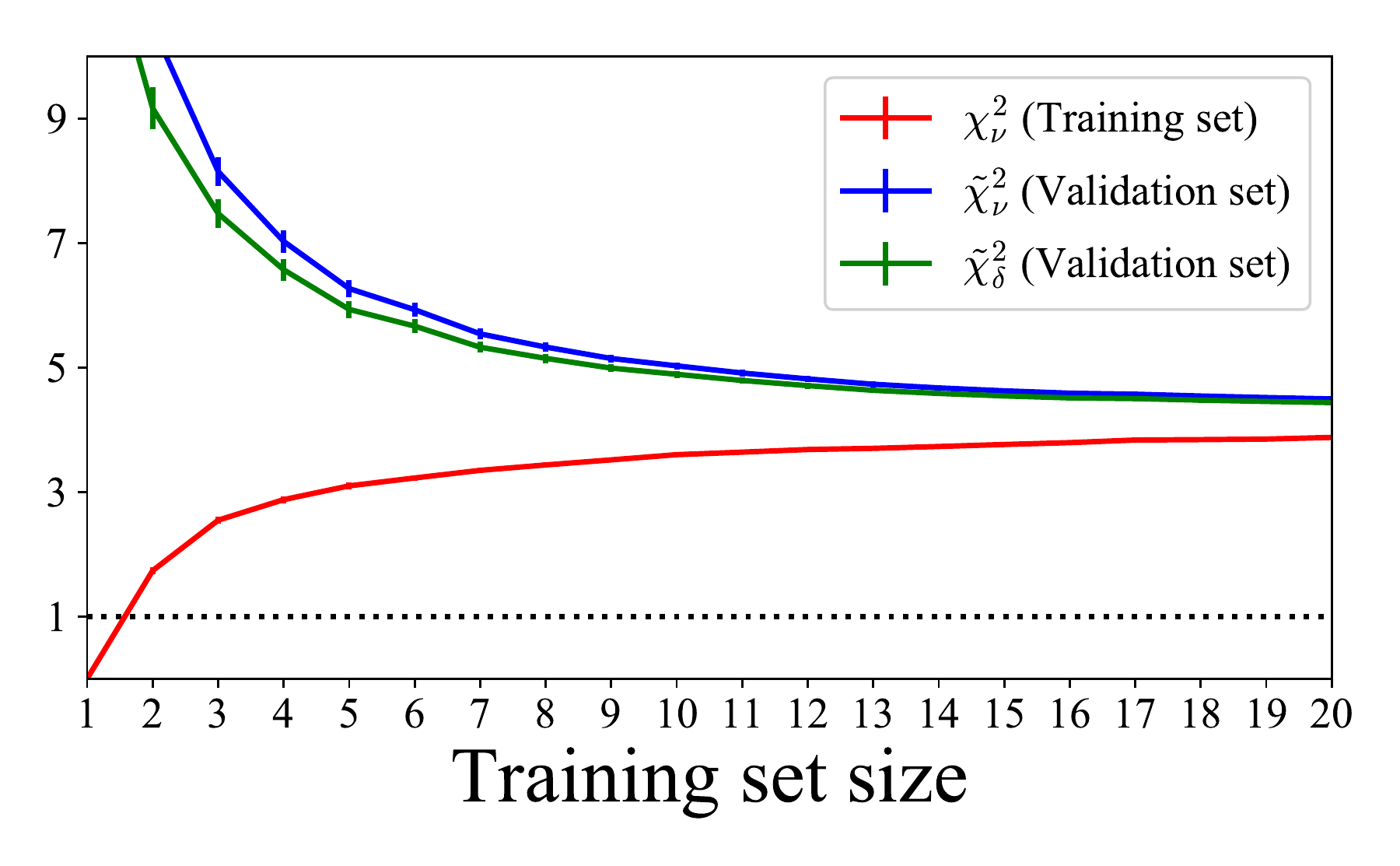} 
\vskip-0.03cm
\includegraphics[width=\columnwidth, trim ={0 2.42cm 0 0.75cm}, clip]{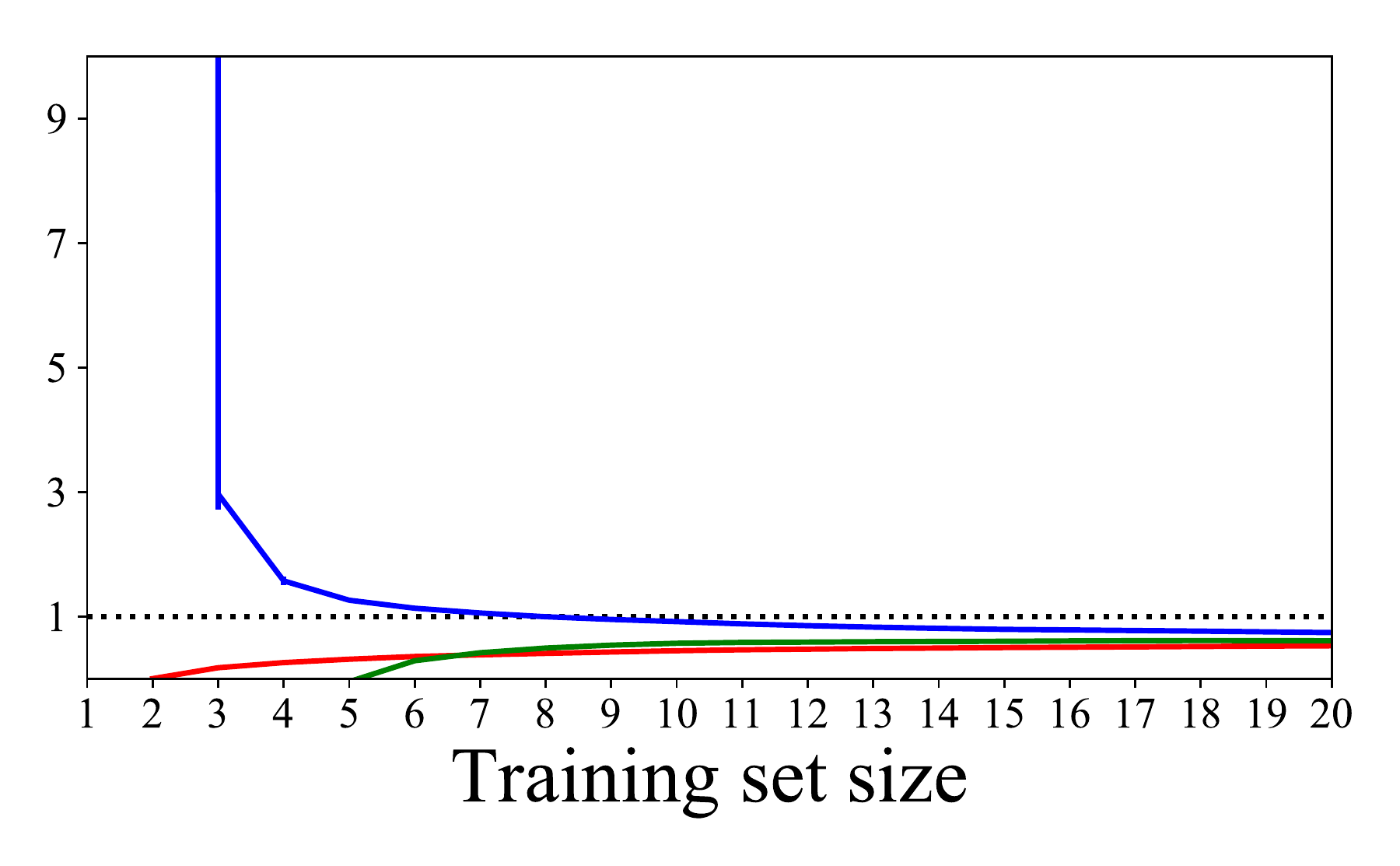} 
\vskip-0.04cm
\includegraphics[width=\columnwidth, trim ={0 2.44cm 0 0.75cm}, clip]{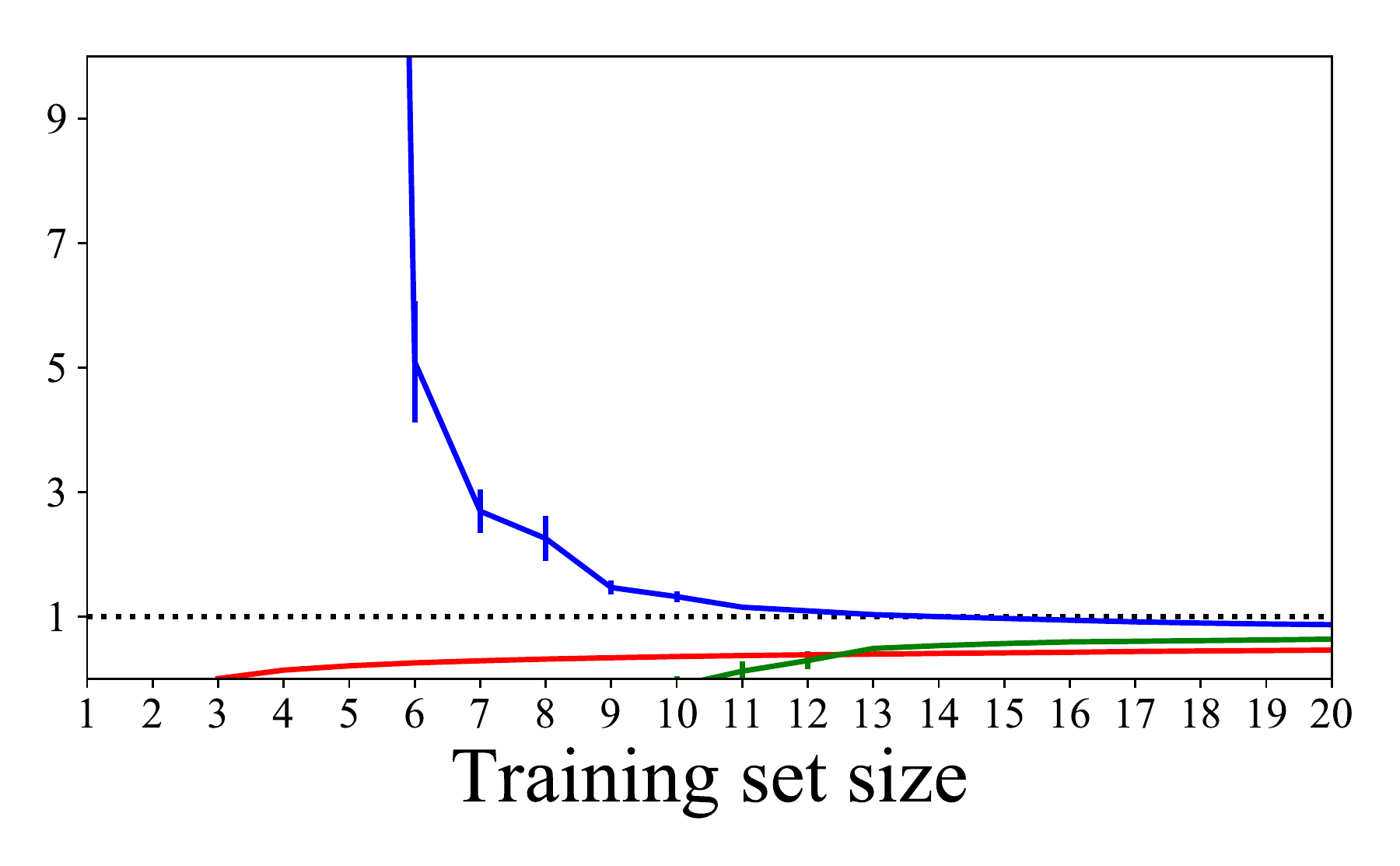} 
\vskip-0.03cm
\includegraphics[width=\columnwidth, trim ={0 0 0 0.75cm}, clip]{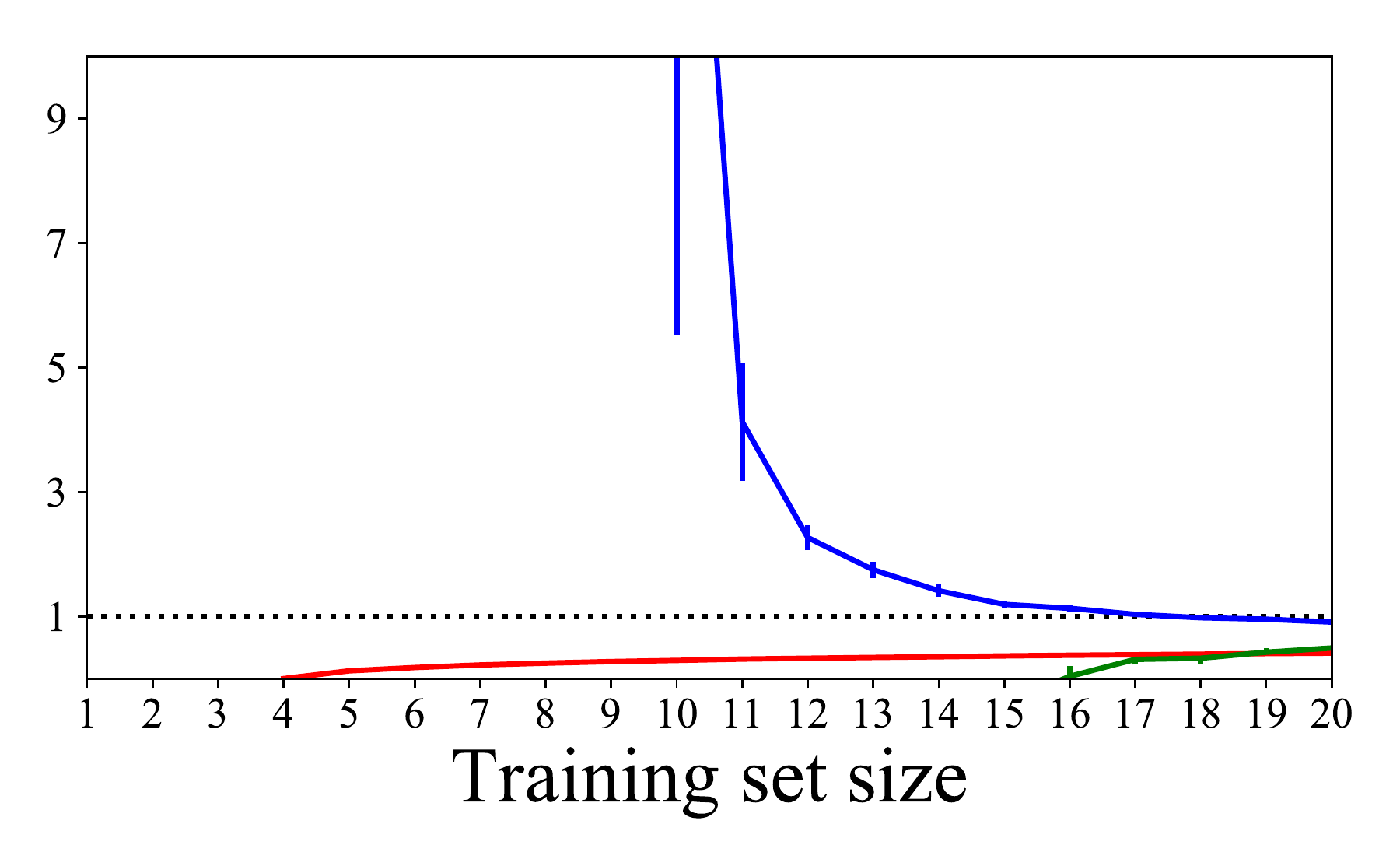} 
\caption{Learning curve analysis for the CC data with $\alpha_{\rm max} = 0, 1, 2, 3$ from top to bottom.}
\label{figlc_cc}
\end{figure}

For the cosmic chronometers, we divide the 31 data points in a training set with $N$ up to 20 and a validation set with $\tilde{N}=11$. Figure \ref{figlc_cc}  shows the learning curves obtained with the template of equation~\eqref{temph} with $\alpha_{\rm max}=\{0,1,2,3\}$ (top to bottom).

The case $\alpha_{\rm max}=0$ is a clear case of under-fitting and is disfavored by the data: for $N=20$ the $\chi^{2}_{\nu}$ is well outside the corresponding 3$\sigma$ interval of $[0.26, 2.17]$ (relative to the $\chi^{2}$ distribution with 19 degrees of freedom).
This case corresponds to a constant Hubble rate, see equation~\eqref{temph}. For the case $\alpha_{\rm max}=1$, $\chi^{2}_{\nu}$ and $\tilde{\chi}^{2}_{\delta}$ converge with a plateau to a value close to 1 and within the corresponding 3$\sigma$ interval $[0.24, 2.21]$.
According to our criteria, this is an optimal value of $\alpha_{\rm max}$.

The case $\alpha_{\rm max}=2$ is similar to the case $\alpha_{\rm max}=1$: $\chi^2_{\nu}$ and $\tilde{\chi}^2_{\delta}$ converge to the expected value with a plateau. Therefore, this case is also optimal, although $\tilde{\chi}^2_{\delta}$ converges to a value a little higher as compared with $\chi^2_{\nu}$, signaling a minor over-fitting. It is worth pointing out that the training curves of Figure \ref{figlc_cc} feature error bars (relative to the mean) computed, as mentioned before, from 2000 partitions. Therefore, the fact that there is a gap between $\chi^2_{\nu}$ and $\tilde{\chi}^2_{\delta}$ is statistically significant.
It is also interesting to note how the learning curves characterize the models:  the case $\alpha_{\rm max}=1$ is clearly simpler than the case $\alpha_{\rm max}=2$ as less data is necessary to train it (it converges faster).

The last case $\alpha_{\rm max}=3$ shows a lack of convergence with plateau, signaling that the model is too complex to be trained by the data.
Therefore, we conclude that this case is disfavored by the data. Finally, we found that  $\alpha_{\rm max}=1$ and  $\alpha_{\rm max}=2$ are both acceptable.
If we were to use the standard analysis based on the $\chi^2_{\nu}$ of equation \eqref{soda} for the full dataset, we would have obtained the results presented in Table~\ref{sodacc}. According to these results $\alpha_{\rm max}=3$ is also acceptable, while the learning curve analysis disfavors it.

\begin{table}[t]
\begin{tabular}{cccc} 
\hline \hline
$\alpha_{\rm max}$ & $\chi^2_{\nu}$ & $3\sigma$ interval & $\nu$ \\
 \hline 
0 & 4.1 & $[0.37, 1.90]$ & 30\\
1 & 0.57 & $[0.36, 1.92]$ & 29\\
2 & 0.53 & $[0.35, 1.94]$ & 28\\
3 & 0.48 & $[0.34, 1.96]$ & 27\\
\hline\hline
 \end{tabular} \\ 
\caption{Analysis based on the $\chi^2_{\nu}$ for the full CC dataset.}
\label{sodacc}
\end{table}

%%%%%%%%%%%%%%%%%%%%%%%%%
%%%%%%%%%%%%%%%%%%%%%%%%%
\subsubsection{Type Ia Supernovae}

\begin{figure}[t]
\centering
\includegraphics[width=\columnwidth, trim ={0 2.44cm 0 0}, clip]{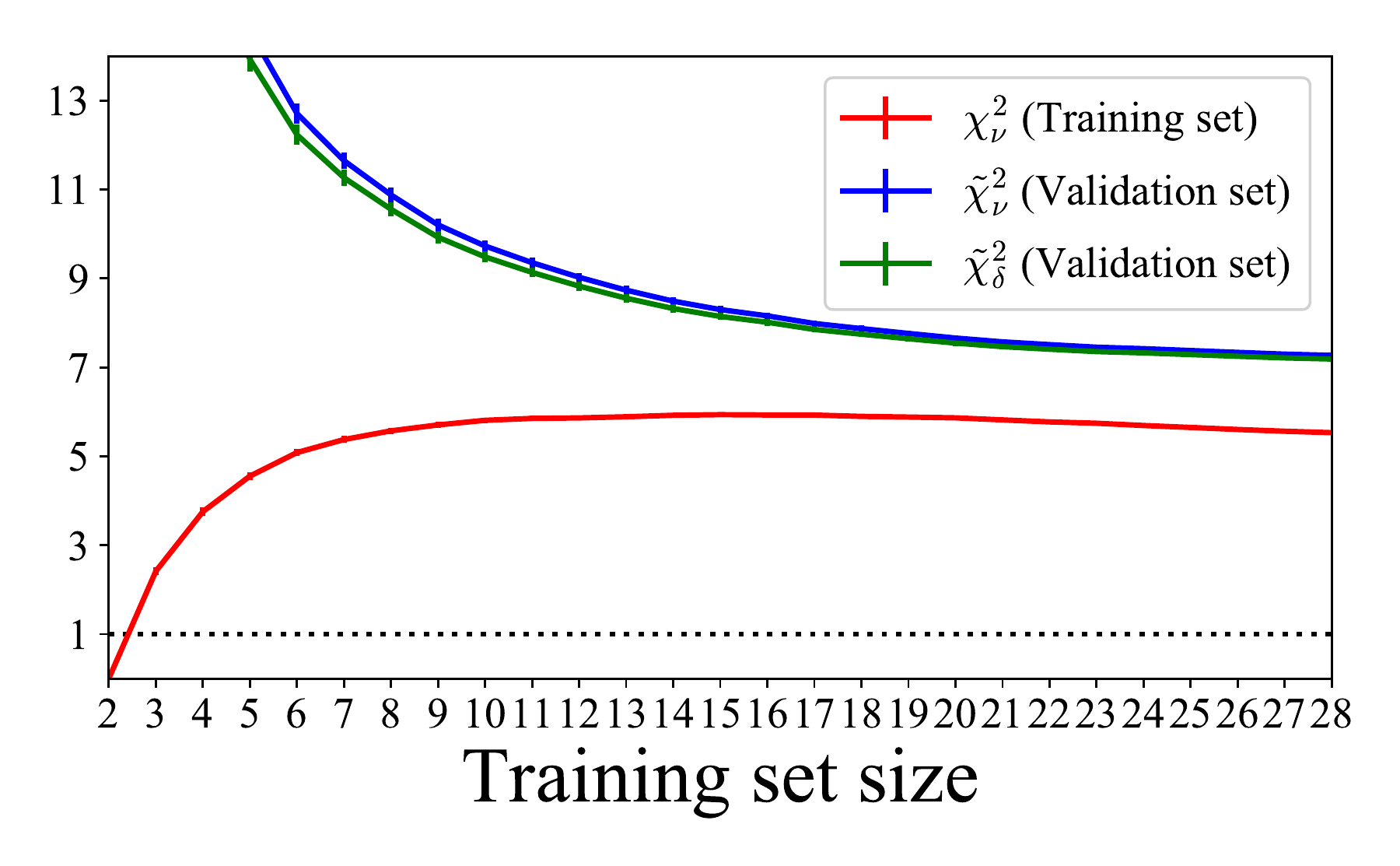} 
\vskip-0.03cm
\includegraphics[width=\columnwidth, trim ={0 2.44cm 0 0.75cm}, clip]{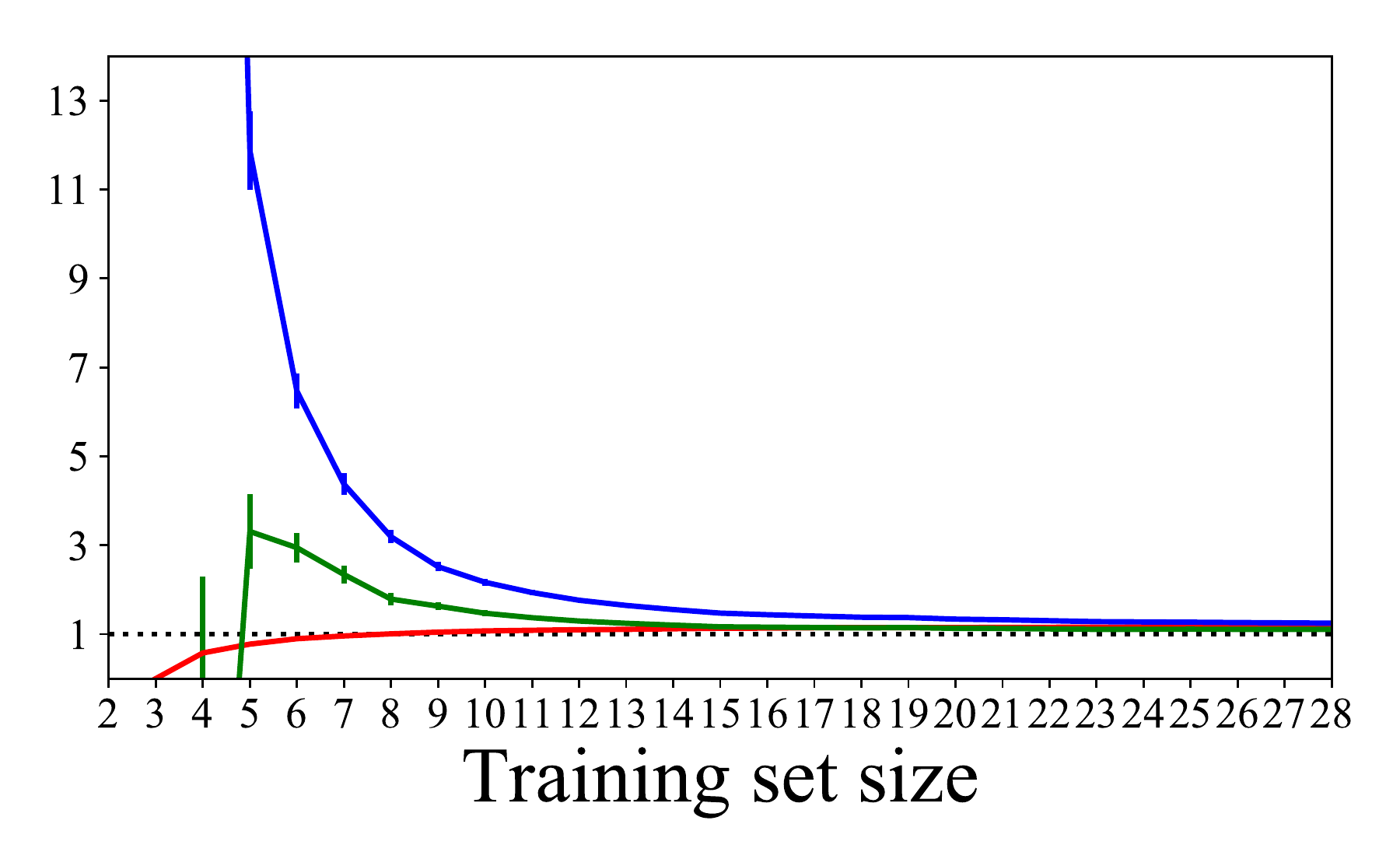} 
\vskip-0.03cm
\includegraphics[width=\columnwidth, trim ={0 2.44cm 0 0.75cm}, clip]{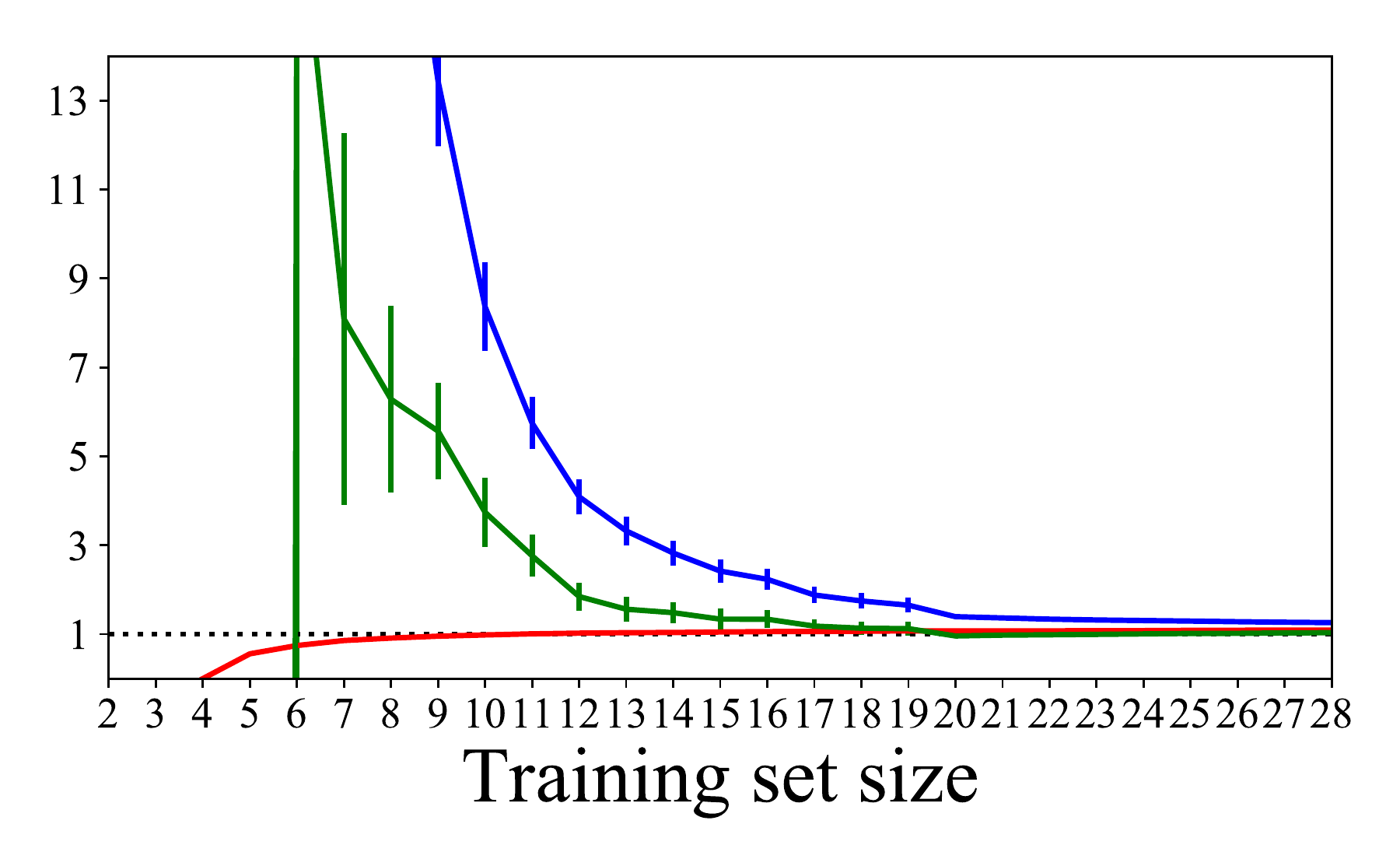} 
\vskip-0.03cm
\includegraphics[width=\columnwidth, trim ={0 0 0 0.75cm}, clip]{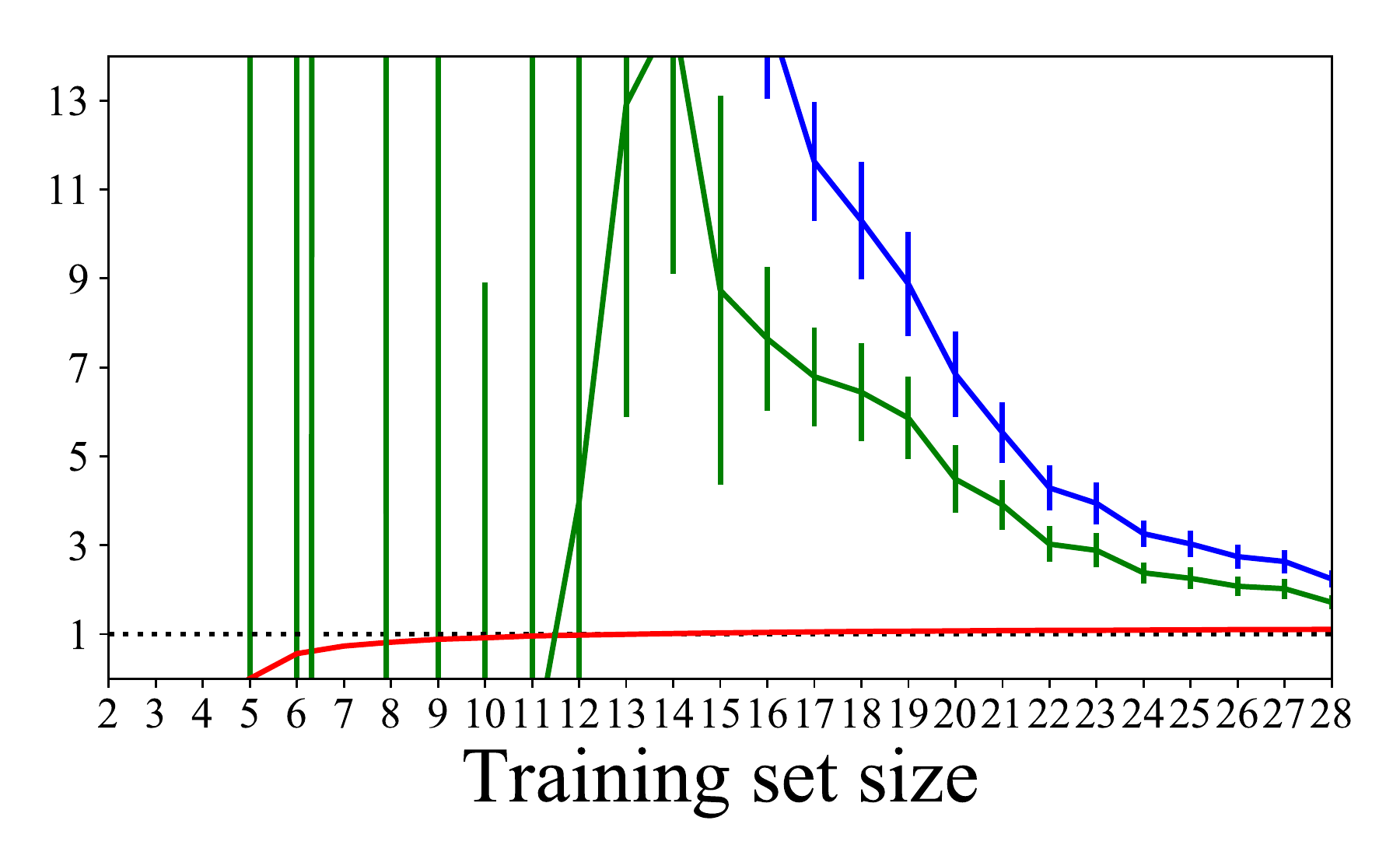} 
\caption{Learning curve analysis for the Pantheon dataset with $\alpha_{\rm max} = 1, 2, 3, 4$ from top to bottom.}
\label{figlc_sn}
\end{figure}

We divide the Pantheon sample of supernovas in a training set with $N$ up to 28 data points and a validation set with $\tilde{N}=12$.
Figure \ref{figlc_sn} shows the learning curves obtained with the template of equation~\eqref{tempmu} with $\alpha_{\rm max}=\{1,2,3,4\}$ (top to bottom). The case $\alpha_{\rm max}=1$ is a clear case of under-fitting and is disfavored by the data. Note that this case coincides with the first order in the cosmographic series expansion~\cite{Cattoen:2007id}. For the case $\alpha_{\rm max}=2$ and $\alpha_{\rm max}=3$, $\chi^{2}_{\nu}$ and $\tilde{\chi}^{2}_{\delta}$ converge with a plateau to a value close to 1 and within the corresponding 3$\sigma$ interval $[0.3, 2.0]$. According to our criteria, these are optimal values of $\alpha_{\rm max}$. Finally, $\alpha_{\rm max}=4$ shows both a lack of convergence and of plateau, signaling over-fitting.

Therefore, we found that  $\alpha_{\rm max}=2$ and  $\alpha_{\rm max}=3$ are both acceptable.
If we were to use the standard analysis based on the $\chi^2_{\nu}$ of equation \eqref{soda} for the full dataset, we would have obtained the results presented in Table~\ref{sodasn}. According to these results $\alpha_{\rm max}=4$ is also acceptable, while the learning curve analysis disfavors it.

%

%%%%%%%%%%%%%%%%%%%%%%%%%
%%%%%%%%%%%%%%%%%%%%%%%%%
\subsubsection{BAO determinations}

\begin{figure}[t]
\centering
\includegraphics[width=\columnwidth, trim ={0 2.44cm 0 0}, clip]{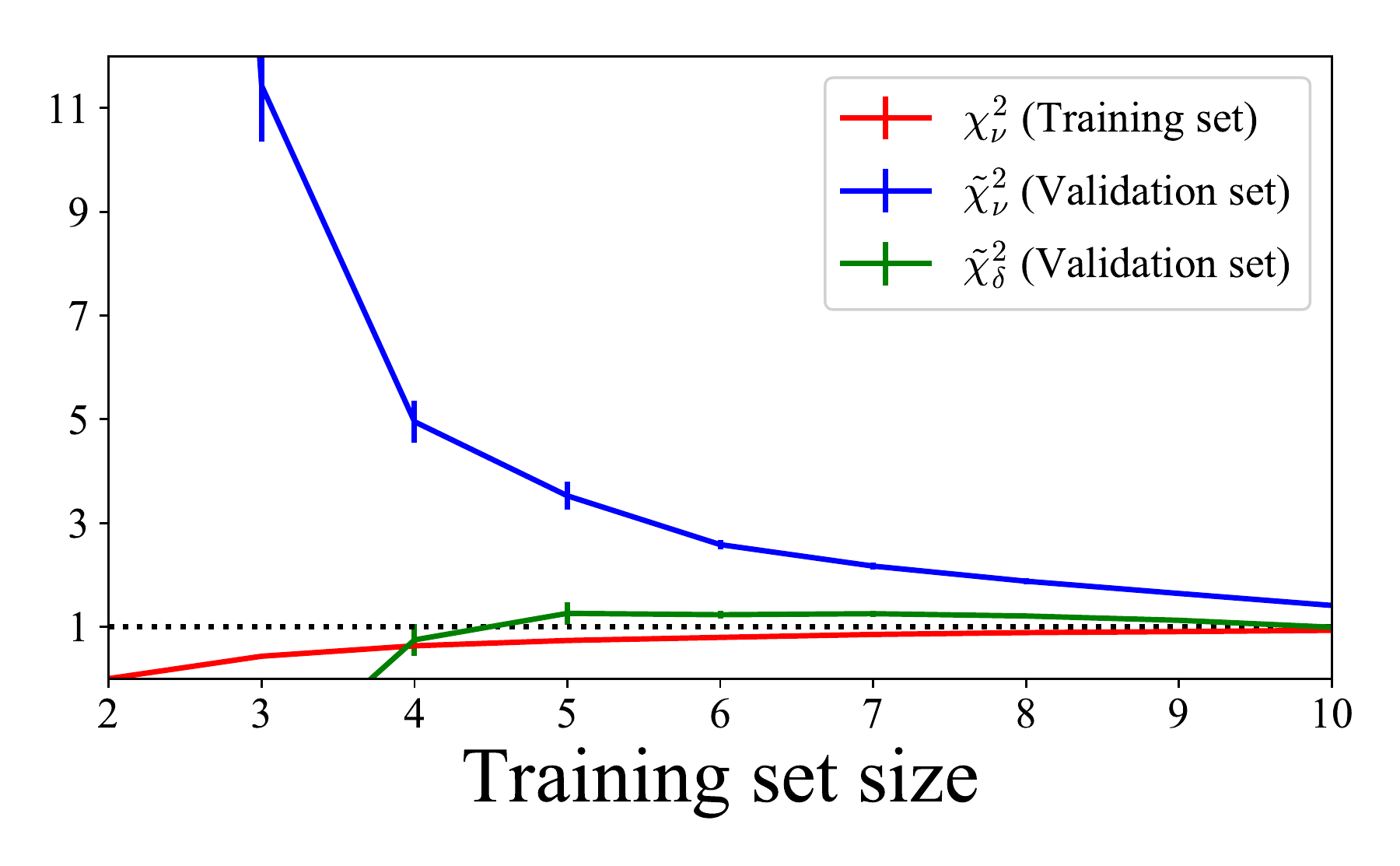} 
\vskip-0.03cm
\includegraphics[width=\columnwidth, trim ={0 0 0 0.75cm}, clip]{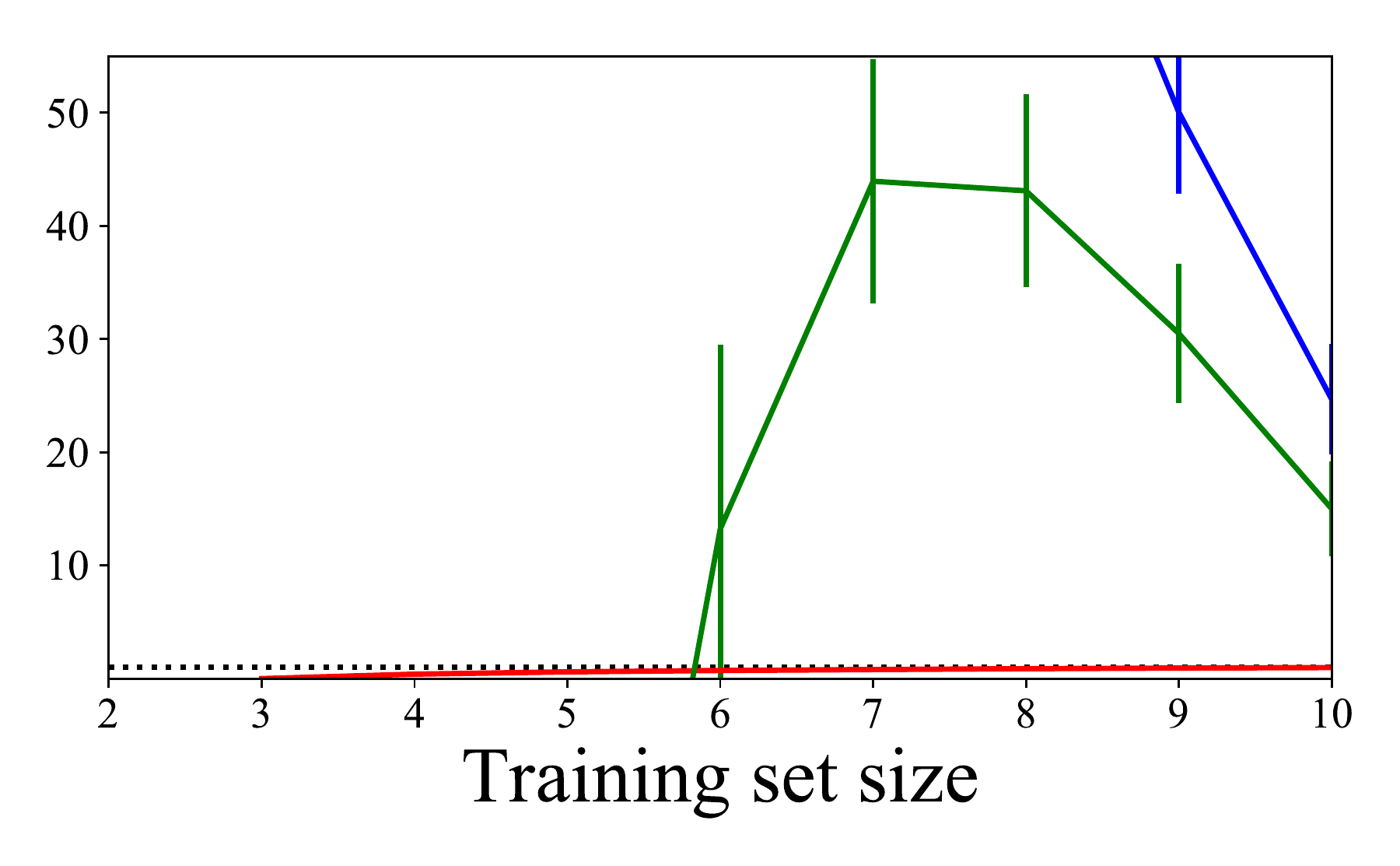} 
\caption{Learning curve analysis for the angular BAO data with $\alpha_{\rm max} = 1$ (top) e $\alpha_{\rm max} = 2$ (bottom).}
\label{figlc_bao}
\end{figure}

With the BAO analysis we divide the 14 data points in a training set with $N$ up to 10 data points and a validation set with $\tilde{N}=4$. Figure \ref{figlc_bao}  shows the learning curves obtained with the template of equation~\eqref{temptheta} with $\alpha_{\rm max}=\{1,2\}$ (top to bottom).

For the case $\alpha_{\rm max}=1$, $\chi^{2}_{\nu}$ and $\tilde{\chi}^{2}_{\delta}$ converge with a plateau to a value close to 1.
According to our criteria, this is the optimal value of $\alpha_{\rm max}$. The case $\alpha_{\rm max}=2$ shows both a lack of convergence and of plateau, signaling over-fitting. We found, therefore, that only  $\alpha_{\rm max}=1$ is acceptable.
If we were to use the standard analysis based on the $\chi^2_{\nu}$ of equation \eqref{soda} for the full dataset, we would have obtained the results presented in Table~\ref{sodabao}. According to these results $\alpha_{\rm max}=2$ is also acceptable, while the learning curve analysis disfavors it.

\begin{table}[h!]
\begin{tabular}{cccc} 
\hline \hline
$\alpha_{\rm max}$ & $\chi^2_{\nu}$ & $3\sigma$ interval & $\nu$ \\
 \hline 
1 & 5.3 & $[0.42, 1.79]$ & 38\\
2 & 1.2 & $[0.42, 1.80]$ & 37\\
3 & 1.1 & $[0.41, 1.81]$ & 36\\
4 & 1.2 & $[0.40, 1.83]$ & 35\\
\hline\hline
 \end{tabular} \\ 
\caption{Analysis based on the $\chi^2_{\nu}$ for the full type Ia SNe dataset.}
\label{sodasn}
\end{table}

\begin{table}[h!]
\begin{tabular}{cccc} 
\hline \hline
$\alpha_{\rm max}$ & $\chi^2_{\nu}$ & $3\sigma$ interval & $\nu$ \\
 \hline 
1 & 1.0 & $[0.14, 2.53]$ & 12\\
2 & 1.1 & $[0.12, 2.62]$ & 11\\
\hline\hline
 \end{tabular} \\ 
\caption{Analysis based on the $\chi^2_{\nu}$ for the full angular BAO dataset.}
\label{sodabao}
\end{table}

%%%%%%%%%%%%%%%%%%%%%%%%%
%%%%%%%%%%%%%%%%%%%%%%%%%
\section{Gaussian Processes}
\label{sec.gp}

A Gaussian Process (GP) is the generalization of the gaussian distribution of a random variable to the infinite function space. This mathematical approach has been successfully used as a nonparametric reconstruction method in cosmology since the pioneering works of \cite{Holsclaw2010,Holsclaw2011}. For instance, it has been applied to different datasets in order to calculate the dark energy equation of state \cite{Holsclaw2010,Holsclaw2011,Seikel2012,Zhang:2018gjb}, the Hubble constant 
\cite{Gomez-Valent:2018hwc,Busti:2014dua,Licia2015,Li2015}, the cosmological matter perturbations \cite{Gonzalez2016,Gonzalez2017b,Yin:2018mvu}, and the gas deplection factor in galaxy clusters \cite{Holanda:2017cmc}, among others.

A GP as a regression method is nonparametric. This means that their predictions are not restricted to a specific functional class (e.g., polinomial), but span an infinite family of classes with properties of continuity and differentiability.
 As this method is based on  Bayesian statistics, we need to use prior and  likelihood distributions  to calculate the posterior distribution.
Both prior and posterior distributions are defined via a mean function and a covariance matrix. The covariance quantifies the correlation between different functional values, $f(z)$ and $f(\tilde{z})$, at arbitrary independent variable points $z$ and~$\tilde{z}$. 

For the prior mean function we adopt the zero function as a conservative choice (this choice is recommended to avoid biased results) and, as commonly used in the literature,  we choose square exponential covariance function:
\begin{equation}\label{eq2}
k(z,\tilde{z})=\sigma_f^2\exp\left(-\frac{(z-\tilde{z})^2}{2l^2}\right).
\end{equation}
The so-called hyperparameters $\sigma_f$ and $l$ are related with the error/variation of the reconstruction and with its  smoothness, respectively. These hyperparameters can be fixed by maximizing the likelihood distribution given the observational data (for a complete description of the GP method see~\cite{Seikel2012,Rasmussen}). To perform the GP regression, we use the python package \texttt{GaPP}.\footnote{Available at \href{http://www.acgc.uct.ac.za/~seikel/GAPP/index.html}{acgc.uct.ac.za/$\sim$seikel/GAPP/index.html}.}

%%%%%%%%%%%%%%%%%%%%%%%%%
%%%%%%%%%%%%%%%%%%%%%%%%%
\section{Results}
\label{sec.res}

Now, we present the reconstructions of $r_{0}(z)$ that we obtained using the methods discussed in the previous sections. The results are divided according to the method and the data used in order to reconstruct $r_{0}(z)$.
%Furthermore, for each method and data, we present two results of the  null test associated to the both priors discussed in section \ref{sec.data}.
In all plots, for comparison  purposes, we include a dotted line corresponding to the value of $r_{0}$ predicted for the $\Lambda$CDM model according to the last results by the Planck satellite~\citep{Aghanim:2018eyx}.
%%%%%%%%%%%%%%%%%%%%%%%%%
%%%%%%%%%%%%%%%%%%%%%%%%%
\subsection{Linear model results}

\begin{figure}[t]
\centering
\includegraphics[width=\columnwidth, trim ={0 2.4cm 0 0}, clip]{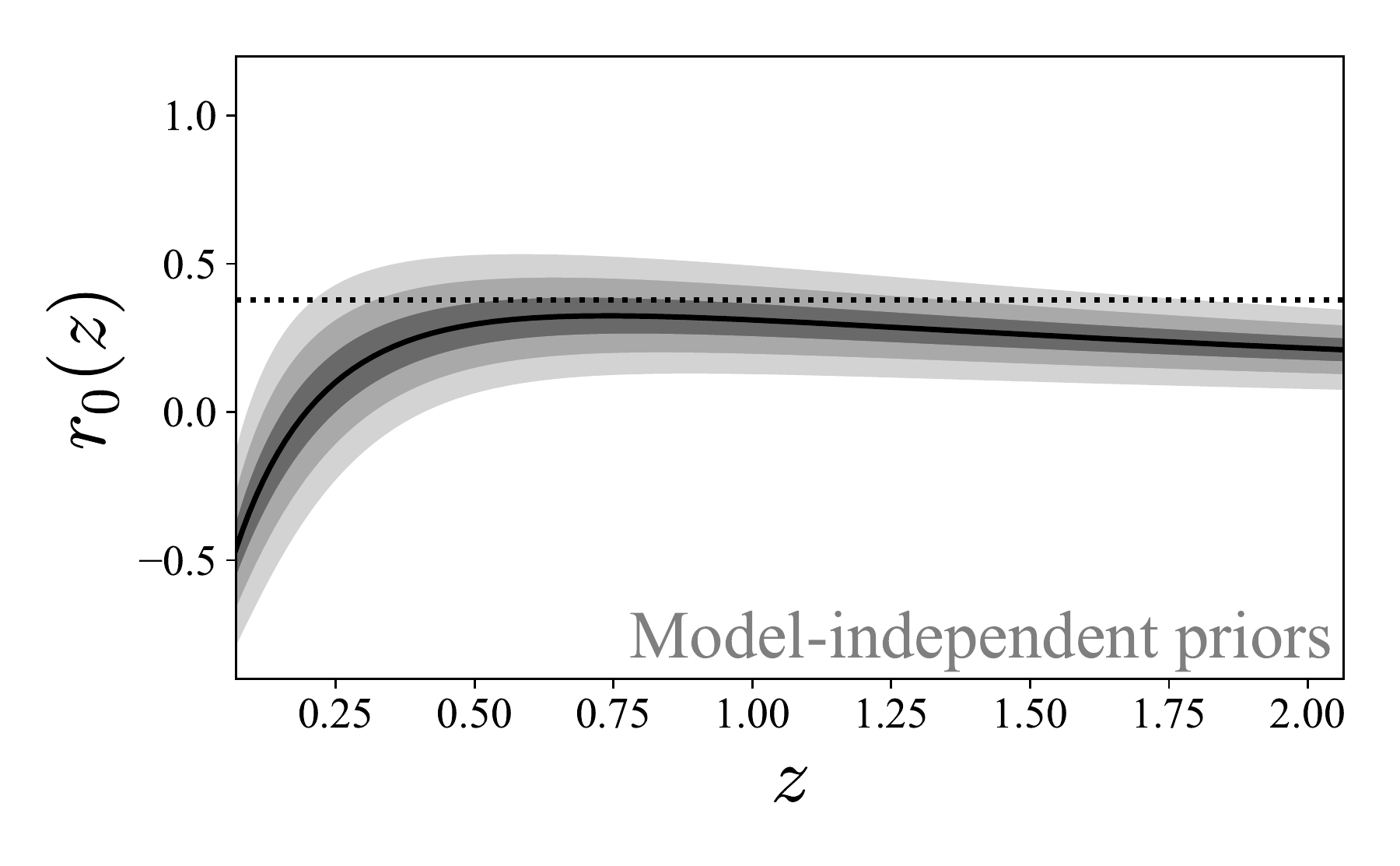} 
\vskip-0.03cm
\includegraphics[width=\columnwidth, trim ={0 0 0 0.75cm}, clip]{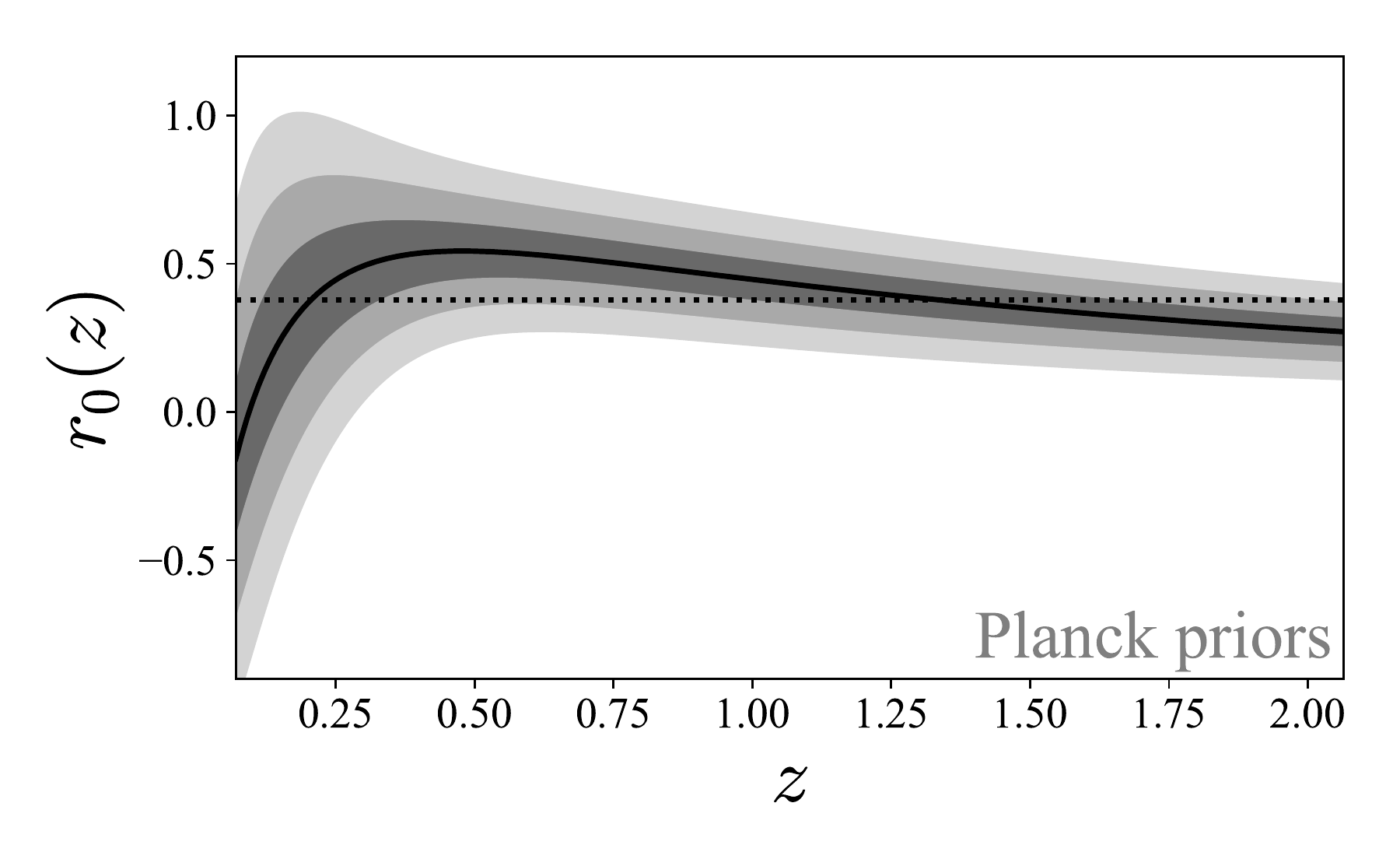}
\vskip-0.3cm
\includegraphics[width=\columnwidth, trim ={0 2.4cm 0 0}, clip]{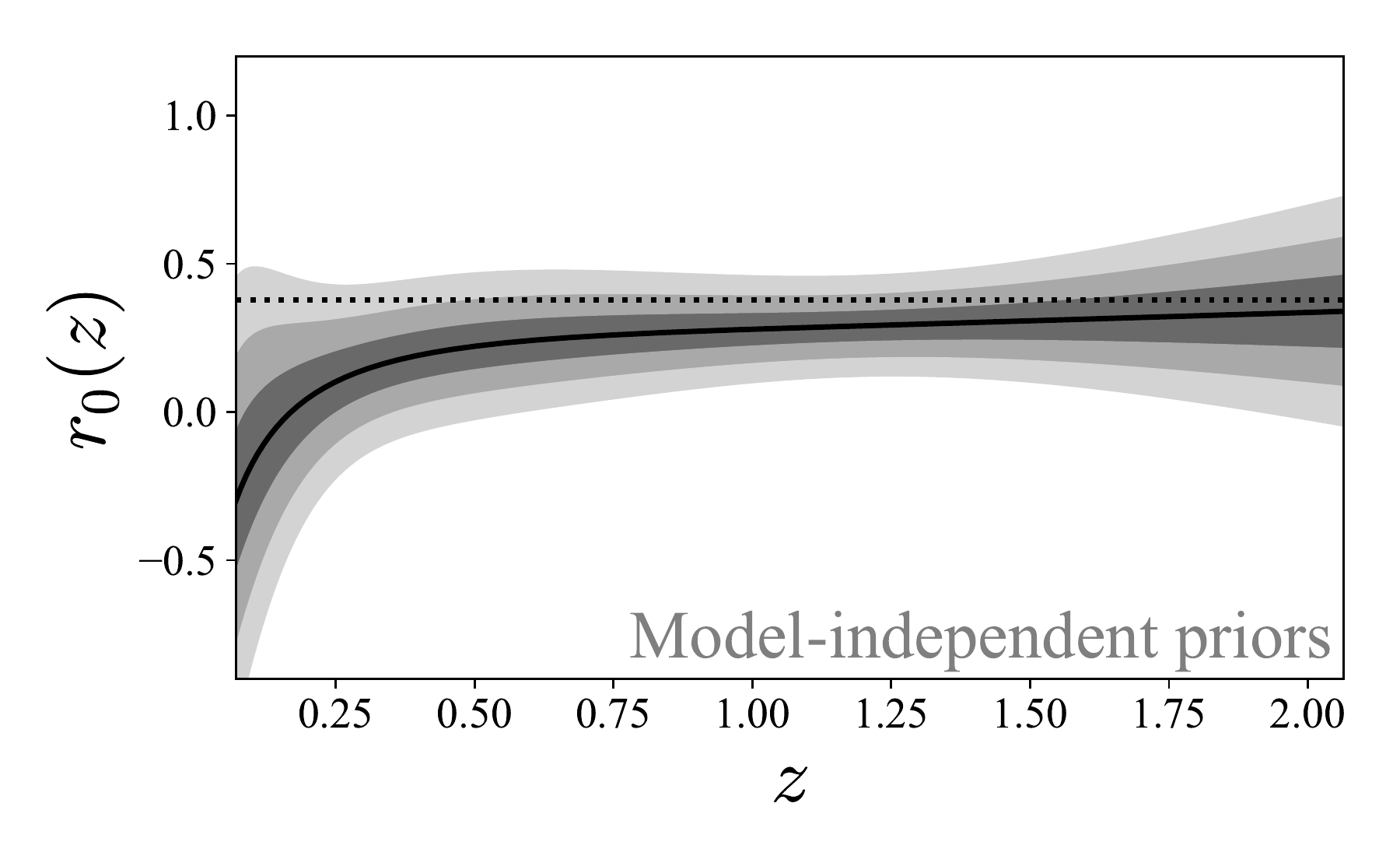} 
\vskip-0.03cm
\includegraphics[width=\columnwidth, trim ={0 0 0 0.75cm}, clip]{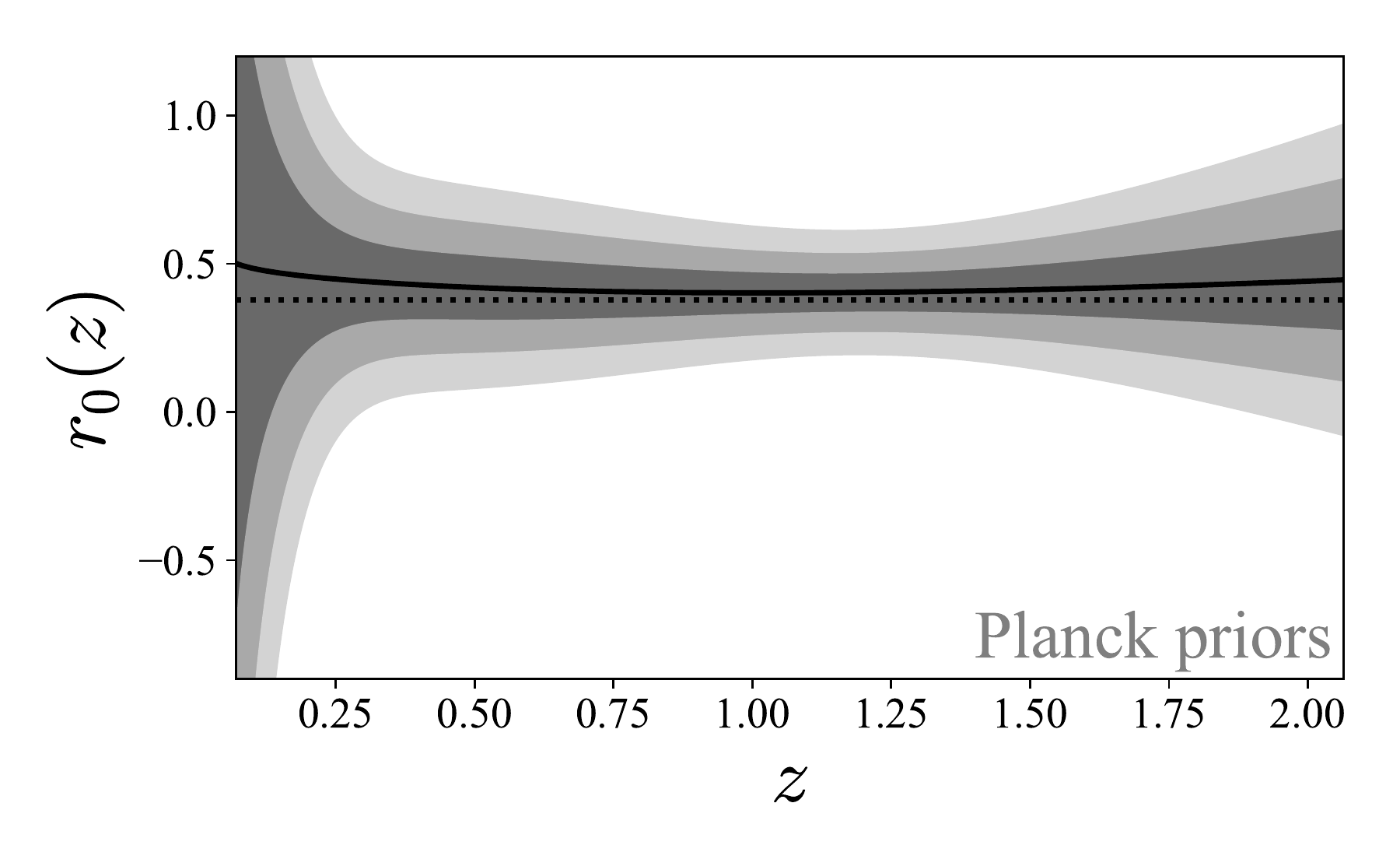} 
\caption{$r_{0}(z)$ obtained from CC data using a LM with $\alpha_{\rm max}=1$ (top) and $\alpha_{\rm max}=2$ (bottom).}
\label{figr01_cc}
\end{figure}

Using the formalism developed in Section~\ref{sec:lin}, we obtain the following results:

\paragraph{\textbf{Cosmic chronometers:}}
Figure~\ref{figr01_cc} shows $r_{0}(z)$ using in equation \eqref{r0z} the Hubble rate function reconstructed via cosmic chronometer data.
For the case $\alpha_{\rm max}=1$ one detects a deviation from the standard model when the model-independent priors discussed in Section~\ref{subsec.ext} are used. This is clearly caused by the determination of the local Hubble constant \eqref{h0riess} which is known to be in tension with the Planck indirect determination.%
\footnote{See \cite{Marra:2013rba,Camarena:2018nbr} for analyses that considered the effect of cosmic variance on local $H_0$.}
Furthermore, the $r_{0}(z)$ test seems to suggest that there is tension only  for $z \lesssim 0.25$ while at higher redshift the local determination of $H_{0}$ gives a reconstruction in agreement with the value expected from Planck.

However, this interesting results loses its significance when $\alpha_{\rm max}=2$ is adopted, which was also found viable. Therefore, we conclude that better CC data are needed in order to conclude on this low-redshift tension.

\begin{figure}[t]
\centering
\includegraphics[width=\columnwidth, trim ={0 2.4cm 0 0}, clip]{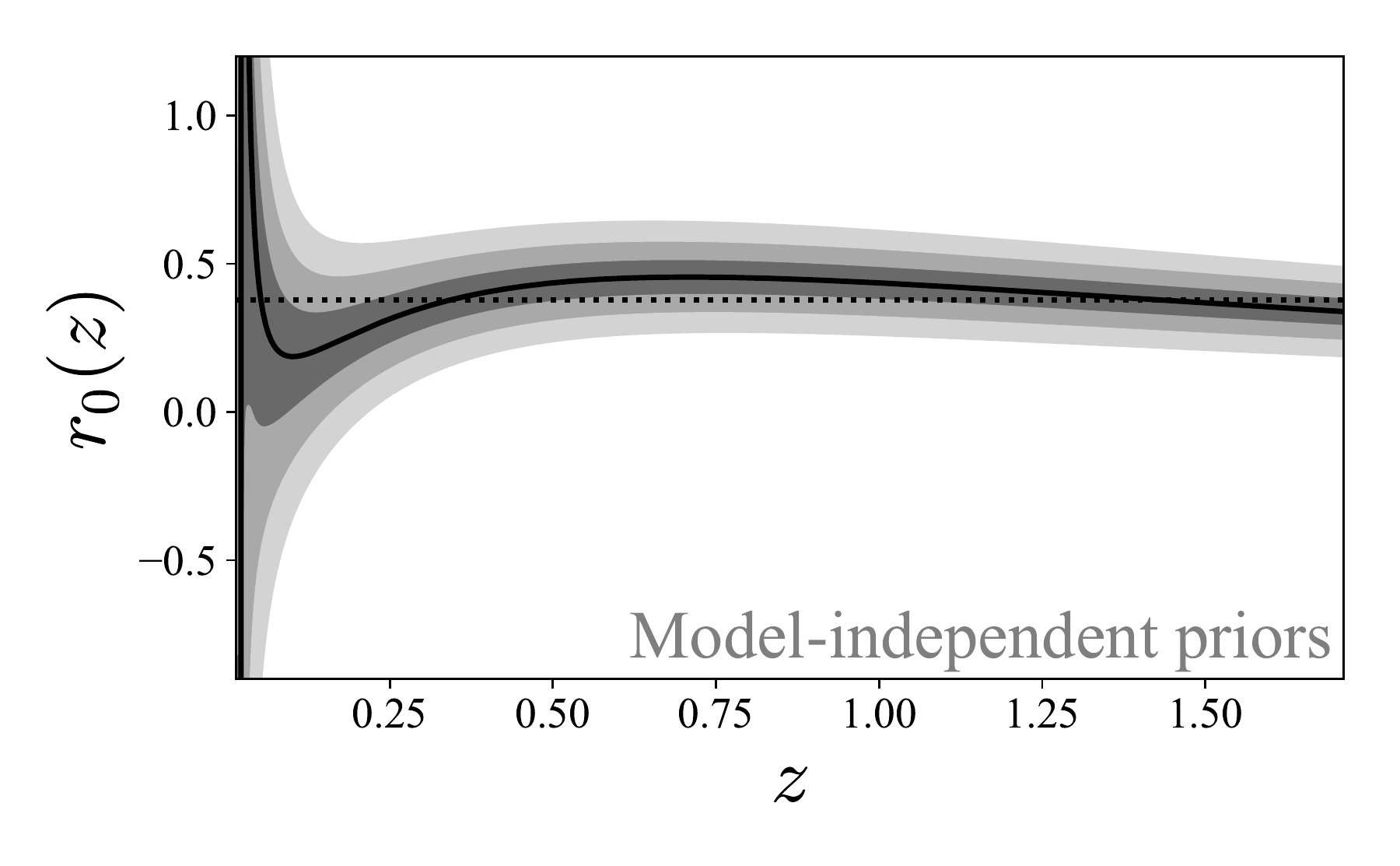} 
\vskip-0.03cm
\includegraphics[width=\columnwidth, trim ={0 0 0 0.75cm}, clip]{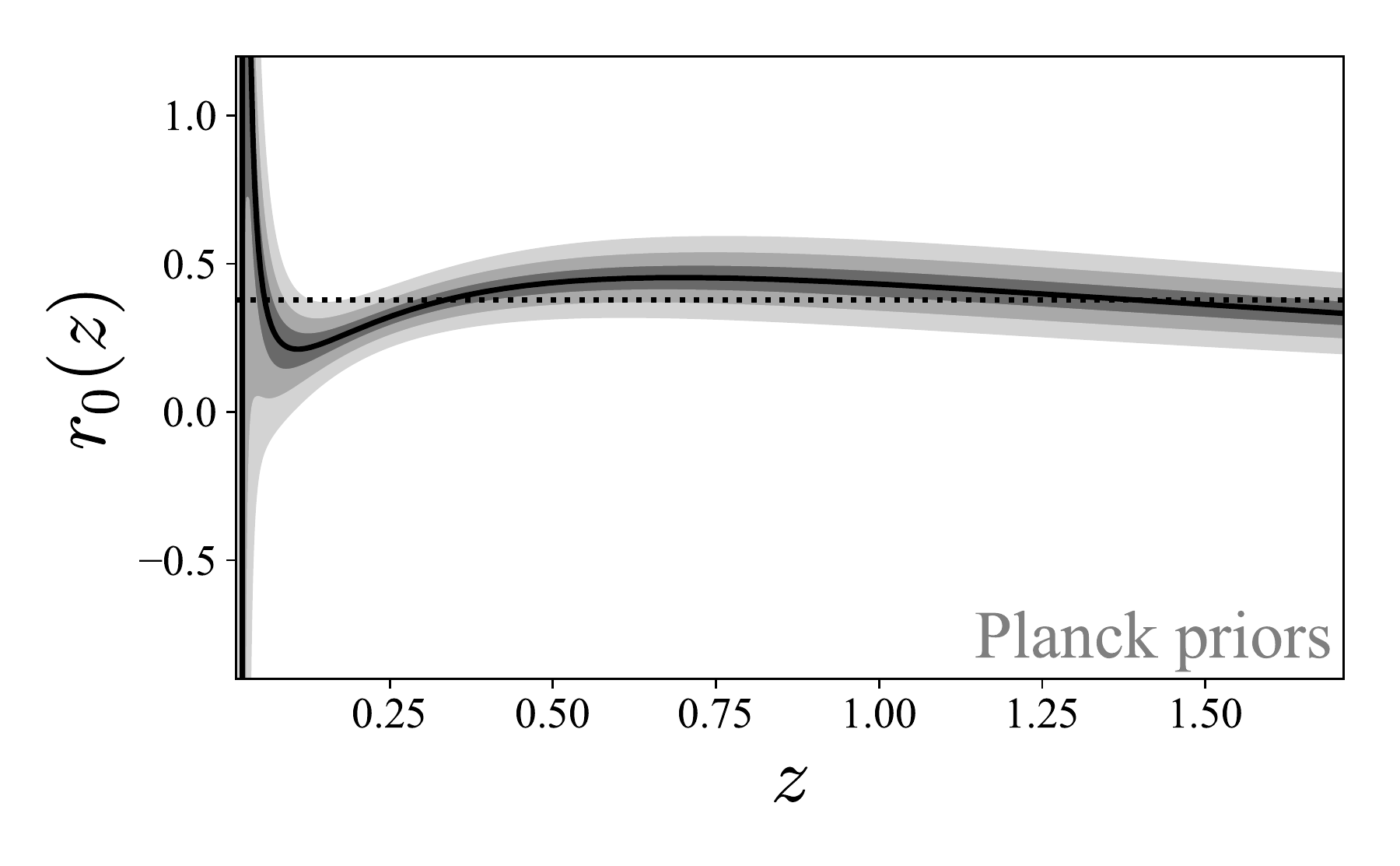} 
\vskip-0.3cm
\includegraphics[width=\columnwidth, trim ={0 2.4cm 0 0}, clip]{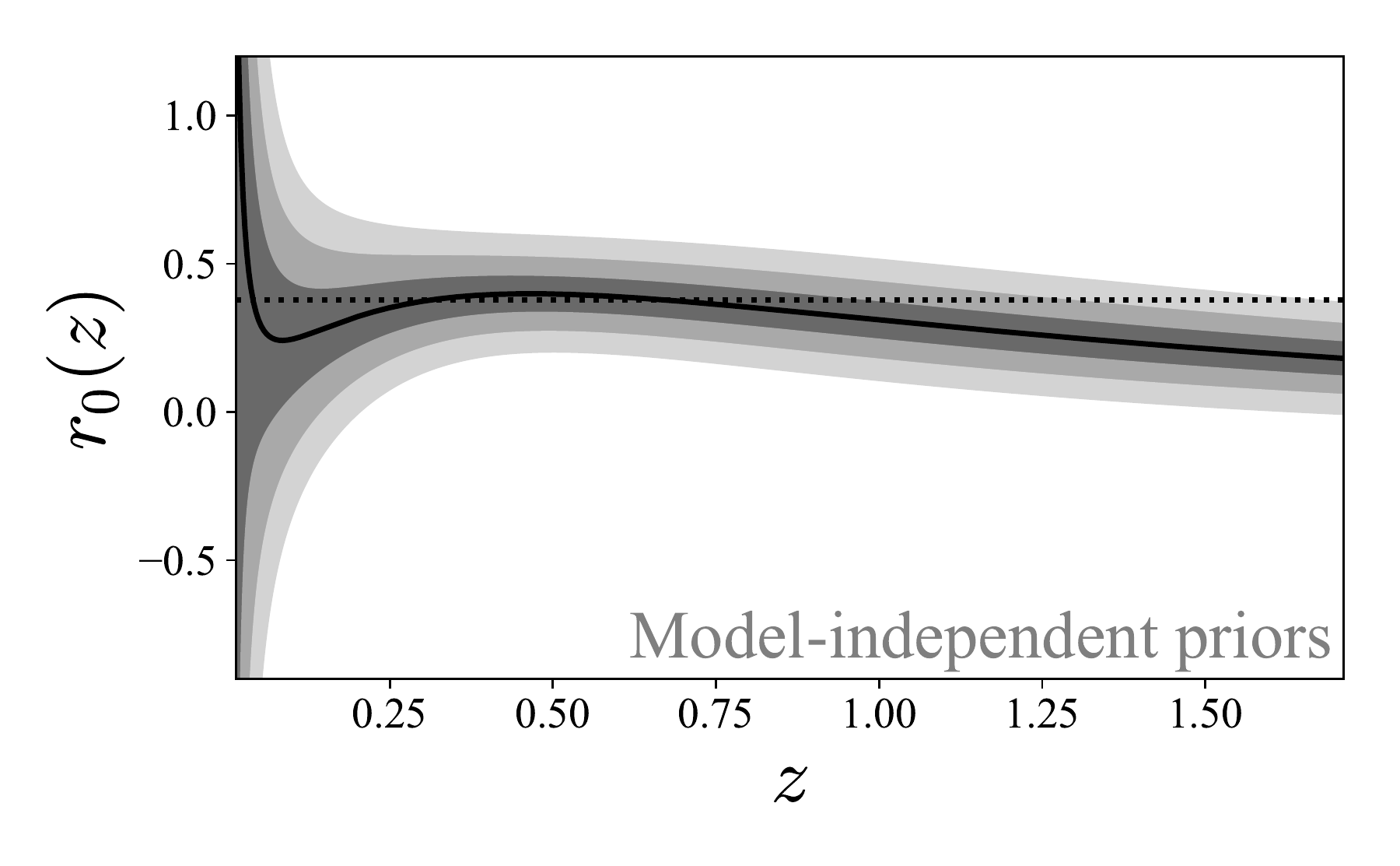} 
\vskip-0.03cm
\includegraphics[width=\columnwidth, trim ={0 0 0 0.75cm}, clip]{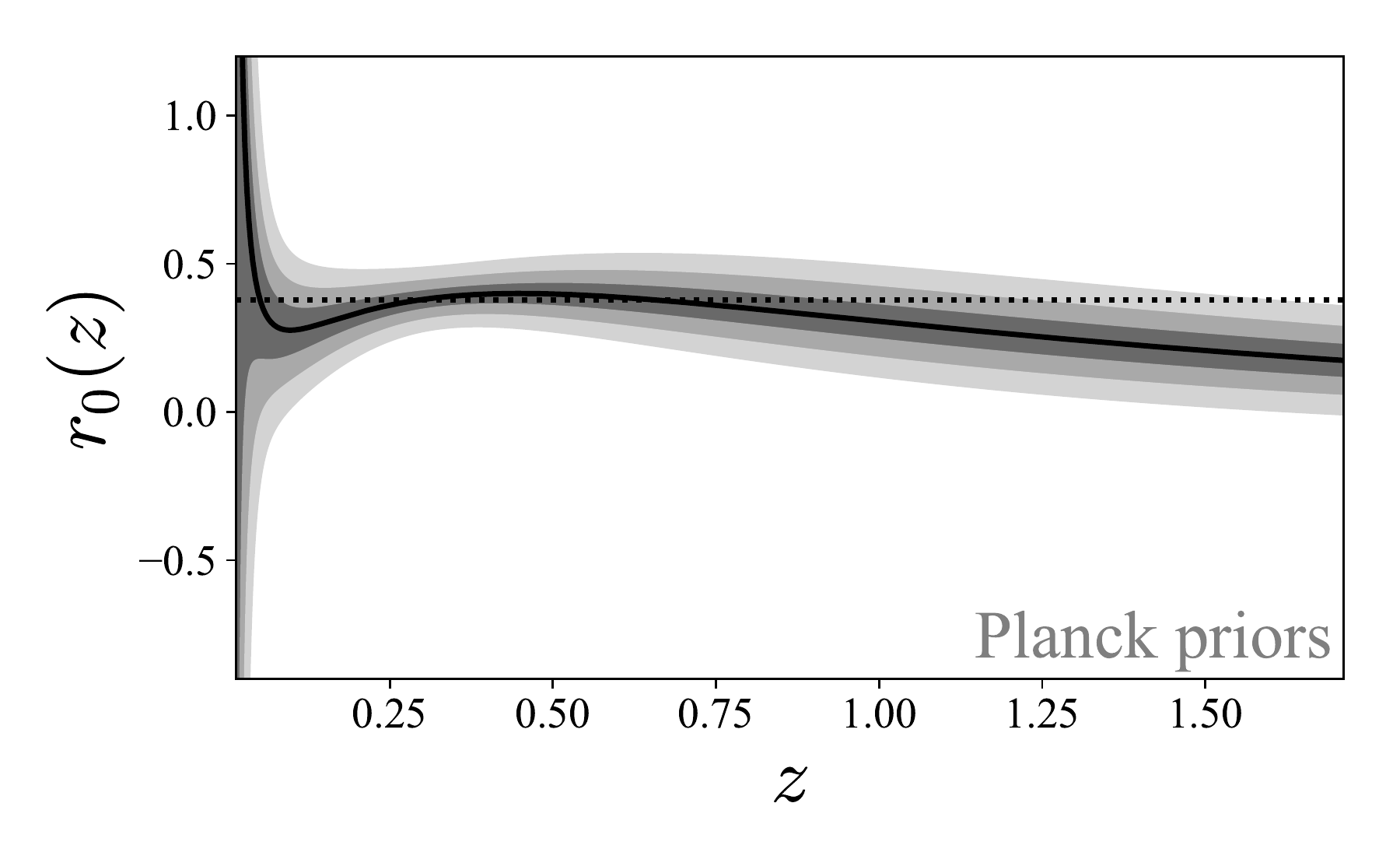} 
\caption{$r_{0}(z)$ obtained from type Ia SN data using a LM with $\alpha_{\rm max}=2$ (top) and $\alpha_{\rm max}=3$ (bottom).}
\label{figr02_sn}
\end{figure}

\paragraph{\textbf{Type Ia Supernovae:}}
Figure \ref{figr02_sn} shows the $r_{0}(z)$ test using the reconstruction of the distance modulus as in equations 
\eqref{Dsn} and \eqref{Dz}.
We do not detect significant deviations from the Planck reference value.

\begin{figure}[t]
\centering
\includegraphics[width=\columnwidth, trim ={0 2.4cm 0 0}, clip]{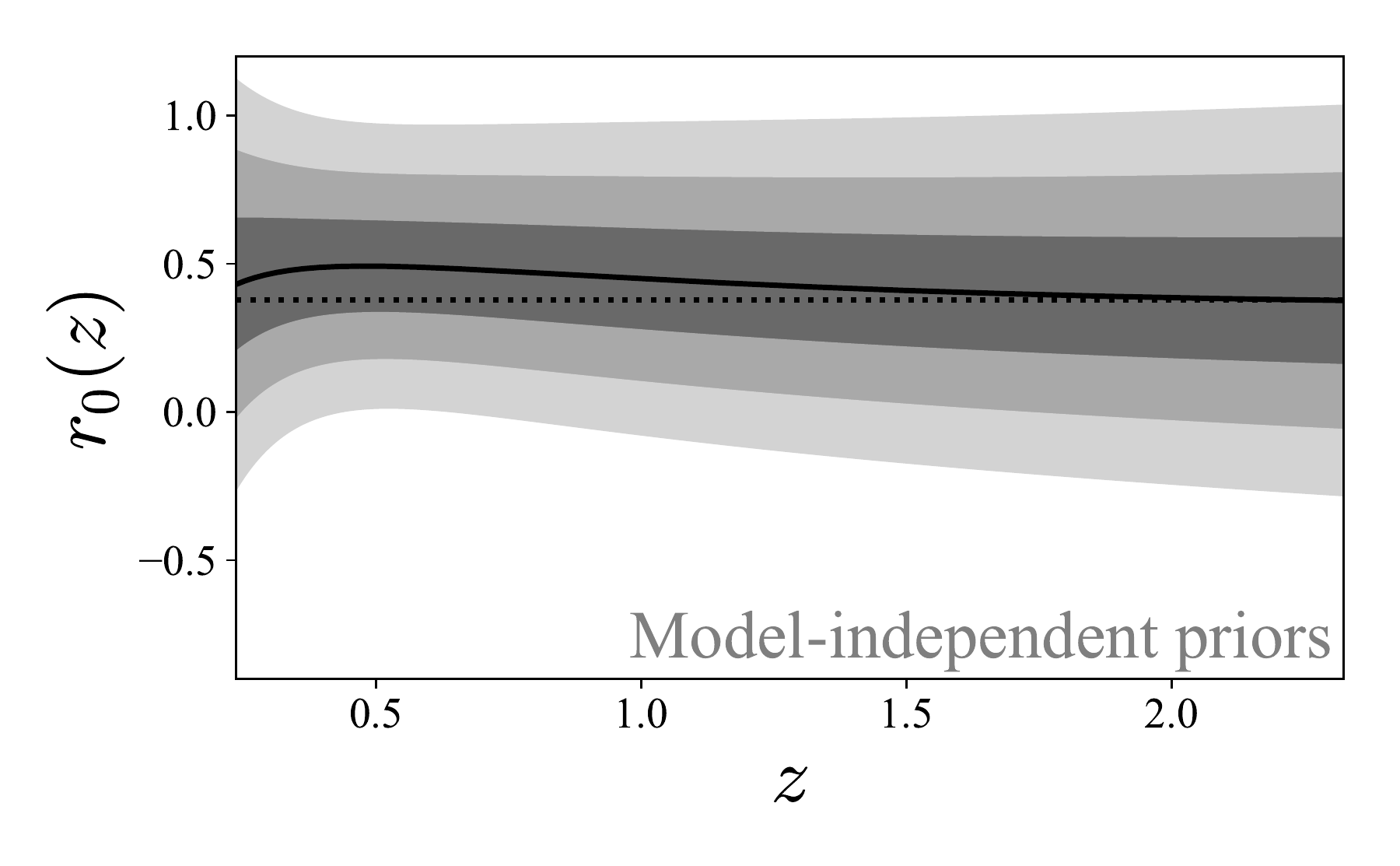} 
\vskip-0.03cm
\includegraphics[width=\columnwidth, trim ={0 0 0 0.75cm}, clip]{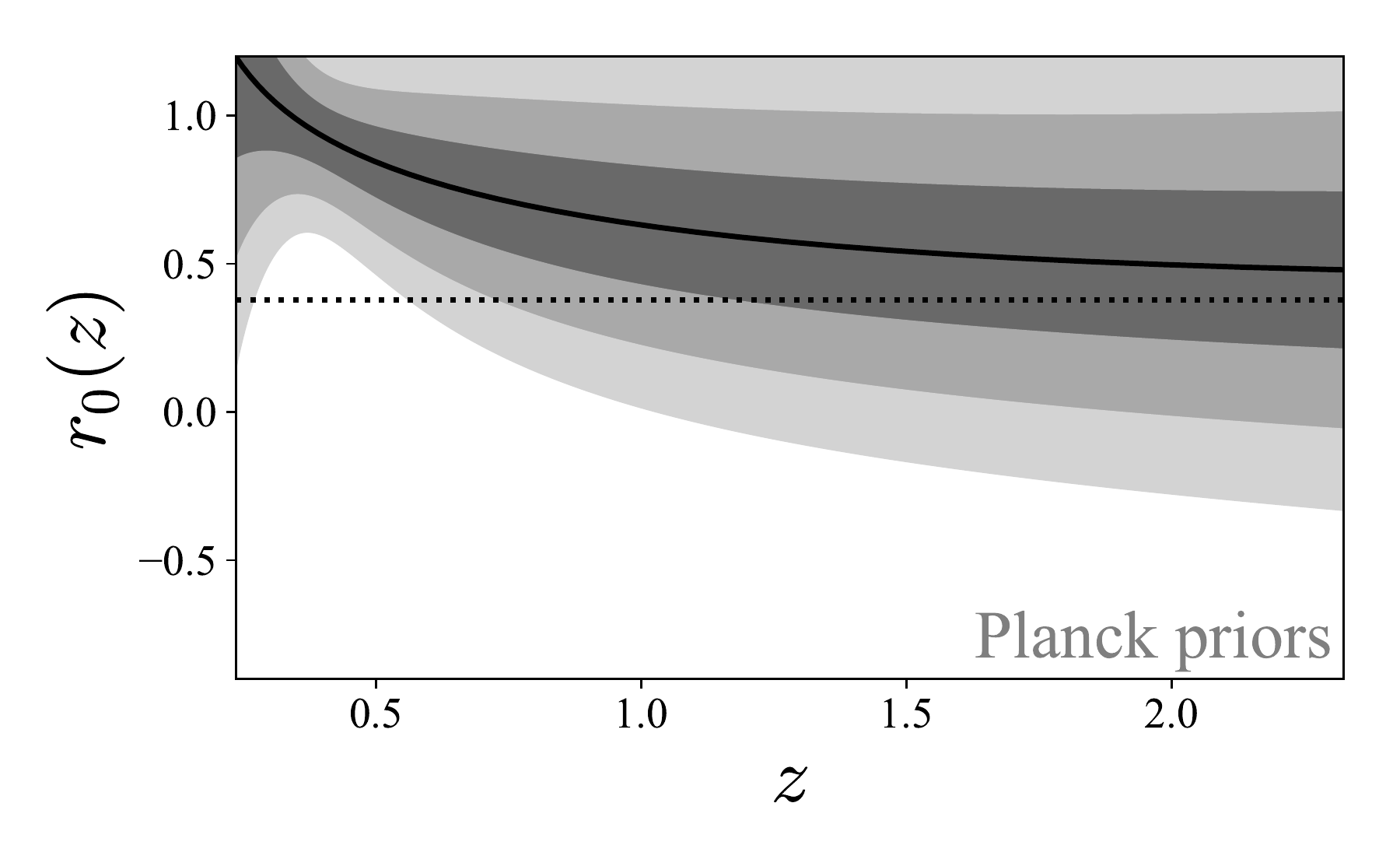} 
\caption{$r_{0}(z)$ obtained from BAO data using a LM with $\alpha_{\rm max}=1$.}
\label{figr01_bao}
\end{figure}

\paragraph{\textbf{BAO:}}
Figure \ref{figr01_bao} shows the $r_{0}(z)$ test using the reconstruction of the BAO angular scale \eqref{theta} from which $E(z)$ is obtained via equations \eqref{Dbao} and \eqref{Dz}.
When using Planck priors we detect a tension at $z\approx 0.3$ with respect to the Planck reference value (but not with a higher constant reference value).
This again suggests a low-redshift deviation from the standard model.

%

%%%%%%%%%%%%%%%%%%%%%%%%%
%%%%%%%%%%%%%%%%%%%%%%%%%
\subsection{Gaussian Process results}
Using the  nonparametric reconstruction of $H(z)$ obtained via GP regression and the $H_0$ and $\omega_b$ priors described in Section~\ref{subsec.ext}, we calculate the null test $r_{0}(z)$ and its confidence levels by Monte Carlo sampling. 

\begin{figure}[t]
\centering
\includegraphics[width=\columnwidth, trim ={0 2.40cm 0 0}, clip]{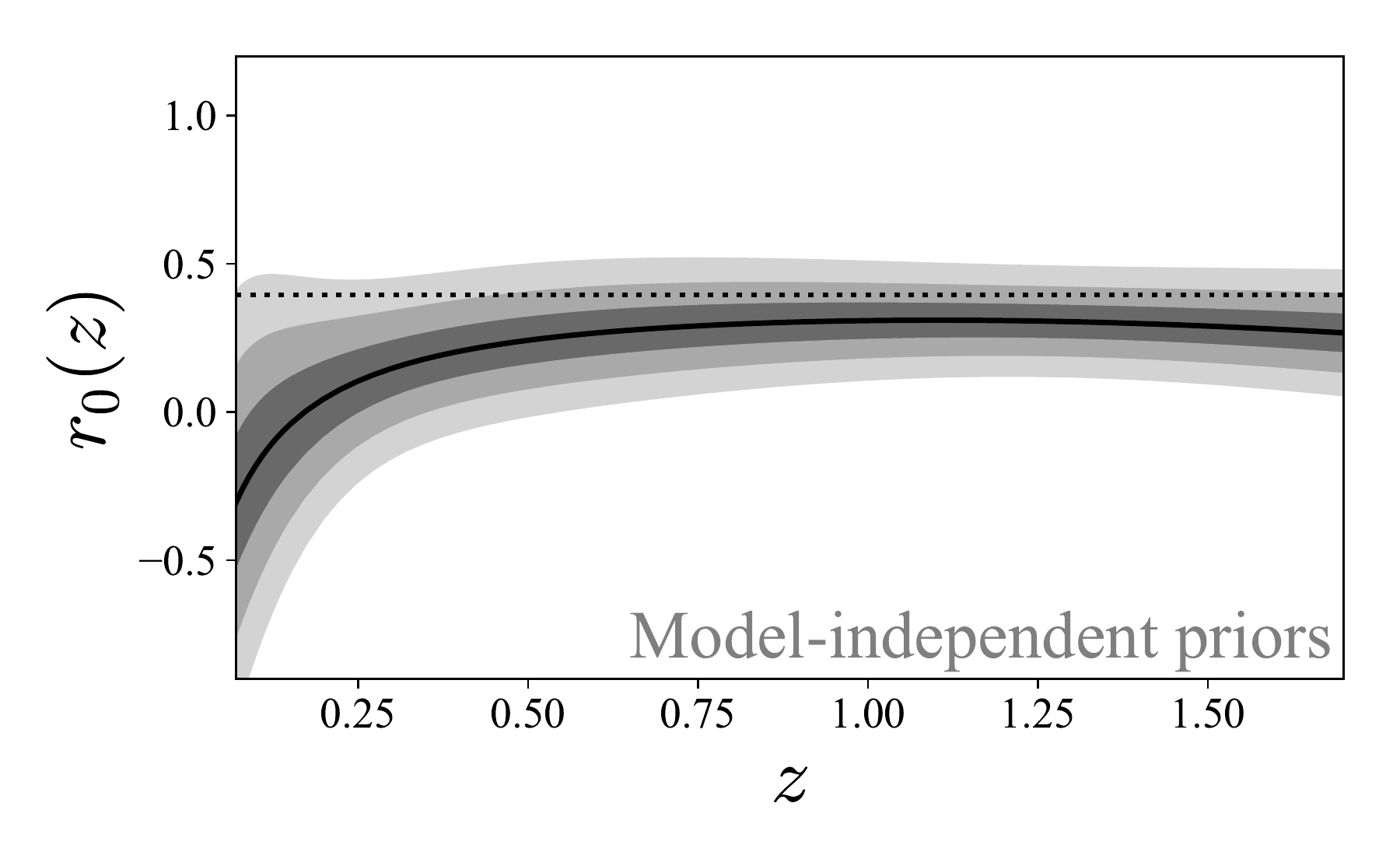} 
\vskip-0.03cm
\includegraphics[width=\columnwidth, trim ={0 0 0 0.75cm}, clip]{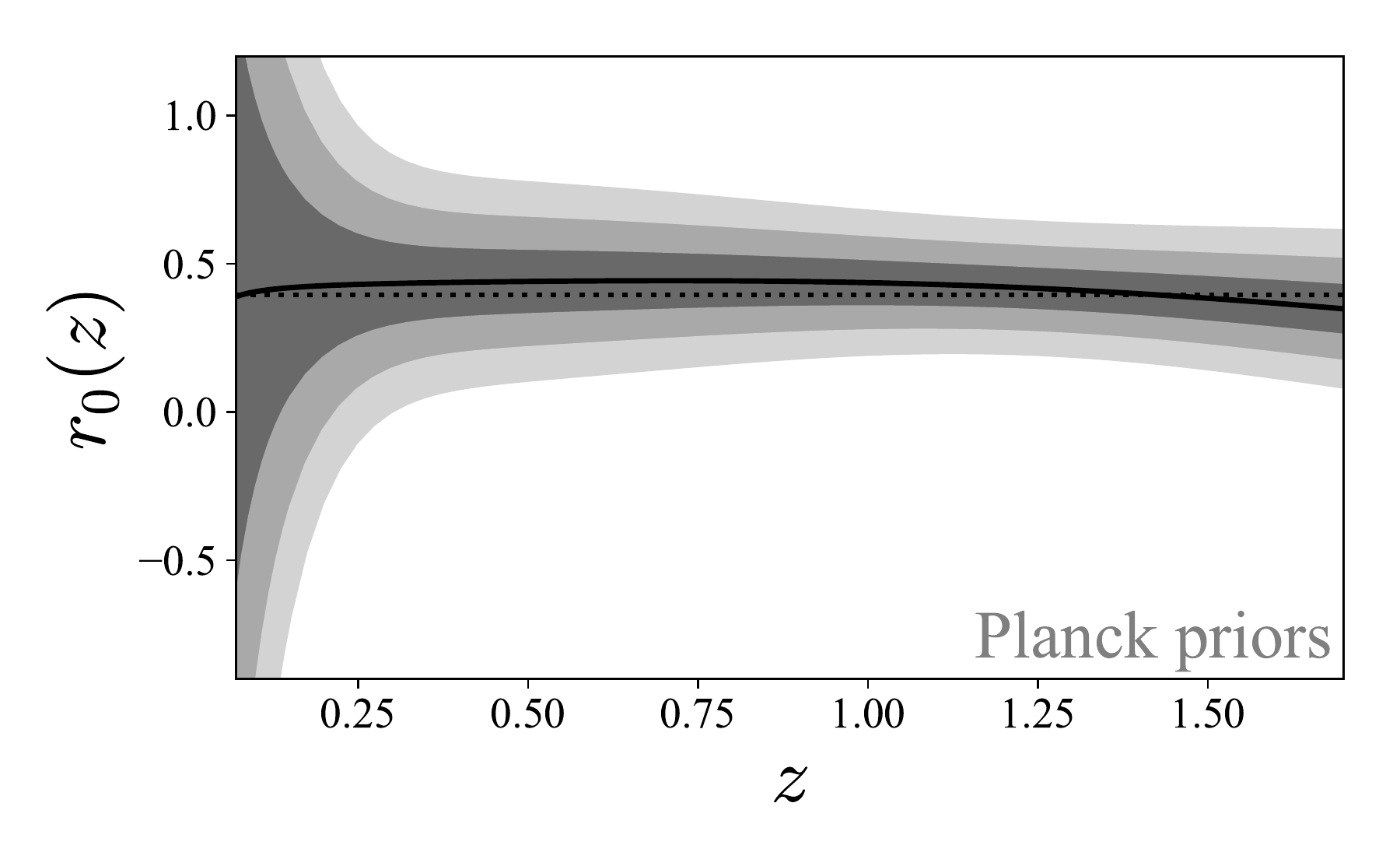} 
\caption{$r_{0}(z)$ obtained from CC data using GPs.}
\label{figr0gp_cc}
\end{figure}

\paragraph{\textbf{Cosmic chronometers:}} In this case, the Hubble rate is reconstructed directly from the CC data which reduces the error propagation and the probability of wiggles in the $H(z)$ reconstruction. Figure \ref{figr0gp_cc} shows the $r_{0}(z)$ calculation using GP method and Monte Carlo sampling. It is evident the similarity of these results for the two sets of $\{H_0,\omega_b \}$ priors with the results obtained using the LM formalism with $\alpha_{\rm max}=2$ (see Figure \ref{figr01_cc}). This emphasizes the possibility that the tension in low-$z$ for the reconstruction via LM with $\alpha_{\rm max}=1$ may be not fundamental.

\begin{figure}[t]
\centering
\includegraphics[width=\columnwidth, trim ={0 2.40cm 0 0}, clip]{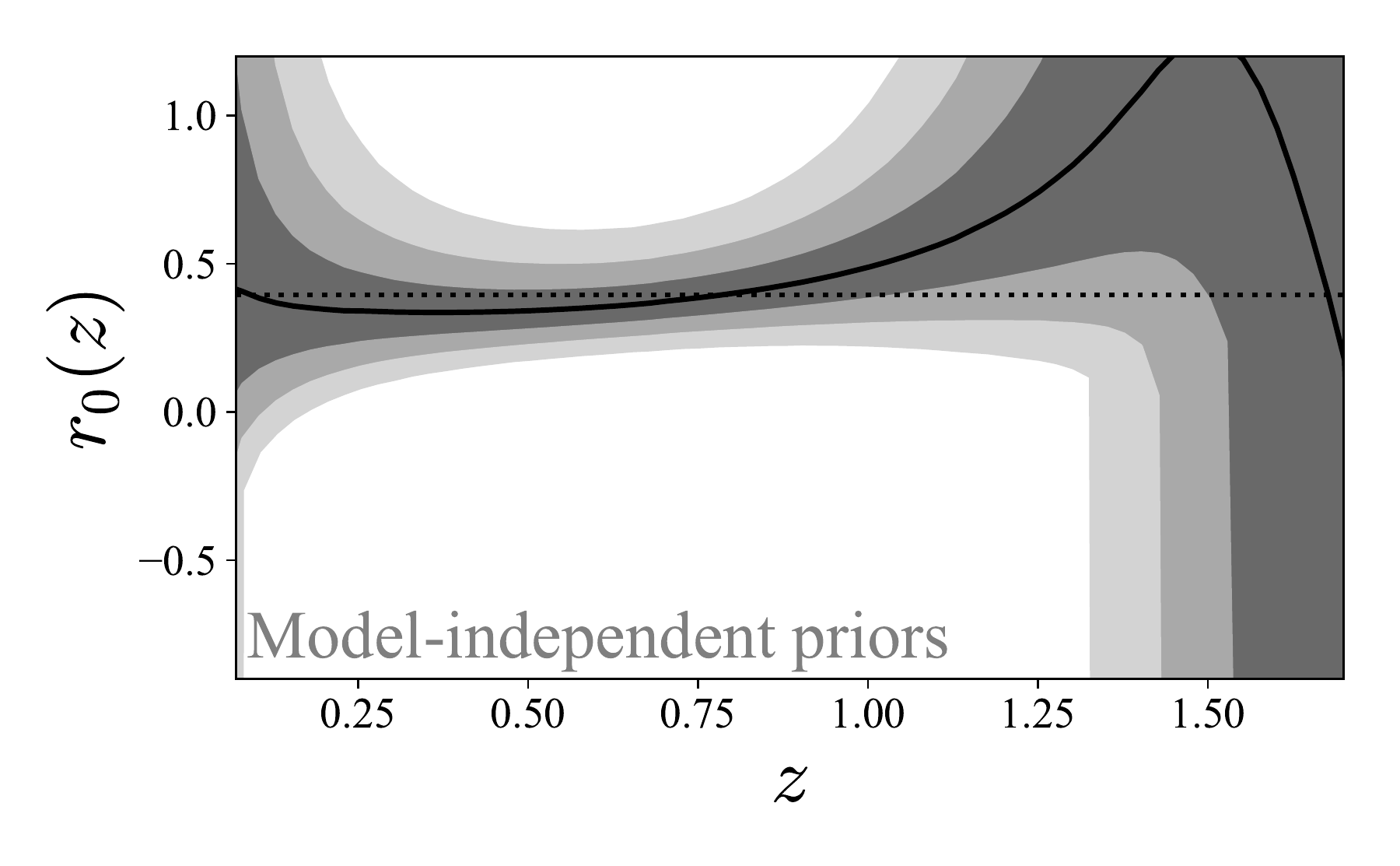} 
\vskip-0.03cm
\includegraphics[width=\columnwidth, trim ={0 0 0 0.75cm}, clip]{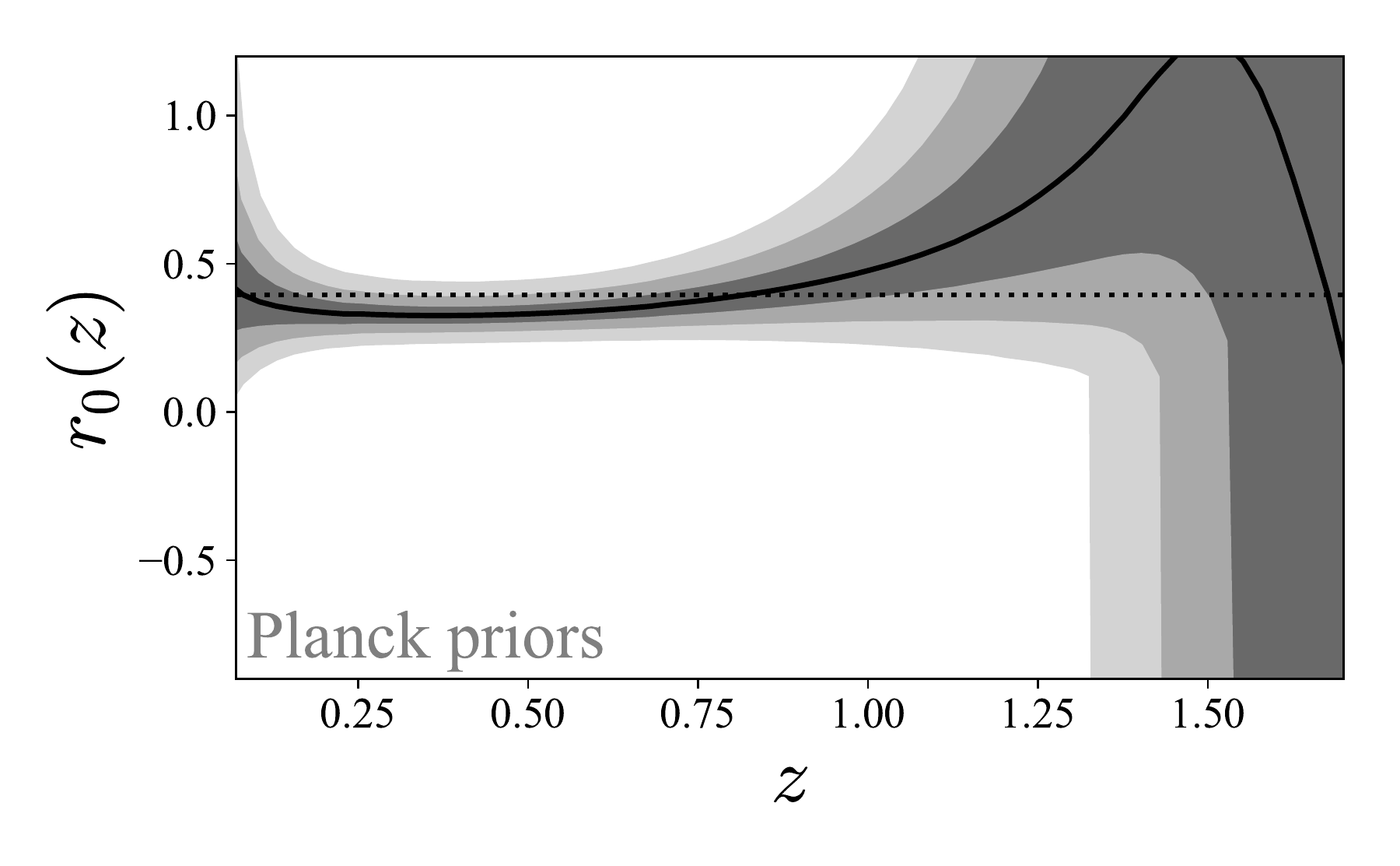} 
\caption{$r_{0}(z)$ obtained from type Ia SNe data using GPs.}
\label{figr0gp_sn}
\end{figure}

\paragraph{\textbf{Type Ia Supernovae:}} Figure \ref{figr0gp_sn} shows  $r_{0}(z)$ from type Ia SN data using GPs. In this case, to calculate the null test we need to transform the distance modulus data to comoving distance and then reconstruct the derivative of this quantity to obtain the Hubble rate (see Section \ref{TIaSn}). The determination of the derivative of ${\mathcal{D}}$ propagates the error and its effect can be seen at high-$z$ where the density of the data is reduced. However, the null test is compatible with the $\Lambda$CDM model in the entire redshift range.

\paragraph{\textbf{BAO:}}
Because of the large gap in redshift between the first thirteen data points and the last one, it is not possible to find a suitable GP reconstruction compatible with a cosmological scenario without assuming a nontrivial prior mean function.

%%%%%%%%%%%%%%%%%%%%%%%%%
%%%%%%%%%%%%%%%%%%%%%%%%%
\section{Conclusions}
\label{sec.conc}

Interacting models of CDM and DE constitute an alternative description of the dark sector which have been largely investigated. In this paper we proposed a new null test in which any deviation of the standard cosmological scenario indicates a non-minimal interaction between these two dark components. Using  linear models and Gaussian processes the expansion rate is reconstructed from the latest CC, SNe Ia and BAO data. For each formalism, the same analysis was performed using two sets of values for the ``external''  parameters $\{H_{0},\Omega_{b},r_{s},M\}$: the first set is based on model-independent results whereas the second one is obtained from the latest results of the Planck collaboration. 

The test performed shows compatibility with the  standard $\Lambda$CDM model within $3\sigma$ confidence level, but some cases deserve a careful analysis. For the LM analysis using the CC data, except the case in which $\alpha_{\rm max}=1$ and the model-independent priors are used, all the other results are compatible with a constant value for $r_{0}\left(z\right)$. The latest Planck result is satisfied in all cases for $z \gtrsim 0.25$. For $z \lesssim 0.25$ only the result with Planck priors reaches satisfactory results. When the model-independent priors are used, there is a considerable tension with the latest Planck result at $z=0$: in the case $\alpha_{\rm max}=1$ there is a severe tension whereas for $\alpha_{\rm max}=2$ there is a $3\sigma$ tension, which is compatible with the current $H_{0}$ tension. For all cases the error range increases when $z \lesssim 0.25$, which means that a possible interaction becoming dynamically relevant at recent times may be a viable possibility.

Still in the LM approach, all the results with SNe Ia data are compatible with a constant value and are in agreement with the latest Planck result. As for the CC result, for $z<0.25$ the result are degenerate. Lastly, for the angular BAO data, the result obtained with model-independent priors is clearly inconclusive since, for all values of $z$, within a $3\sigma$-range all interval $\left[ 0,1 \right]$ is admissible for $r_{0}\left(z\right)$. This degenerate result is related to the big error in the model-independent determination of~$r_{s}$. However, using Planck priors, there is a disagreement signature when $z\approx0.3$. For the GP analysis, all the results are consistent with a constant value and also are in agreement with the $\Lambda$CDM Planck result.

%%%%%%%%%%%%%%%%%%%%%%%%%
%%%%%%%%%%%%%%%%%%%%%%%%%
\begin{acknowledgments}
%It is a pleasure to thank XYZ for useful comments and discussions.
RvM acknowledges support from CNPq and the Federal Commission for Scholarships for Foreign Students for the Swiss Government Excellence Scholarship (ESKAS No. 2018.0443) for the academic year 2018-2019.
VM thanks CNPq and FAPES for partial financial support. JEG acknowledges support from CNPq (PDJ No. 155134/2018-3).
JSA acknowledges  support  from  CNPq  (Grants  no.~310790/2014-0 and 400471/2014-0) and FAPERJ (Grant no.~204282). 
\end{acknowledgments}

%%%%%%%%%%%%%%%%%%%%%%%%%
%%%%%%%%%%%%%%%%%%%%%%%%%
%\bibliographystyle{utphys}
%\bibliography{r0}

\providecommand{\href}[2]{#2}\begingroup\raggedright\endgroup

\appendix

\section{Calibrated MSPE}
\label{conta}

In order to obtain a performance estimator with the same expectation value of $\chi^2_{\nu}$, let us rewrite equation \eqref{co1} as:
\begin{align} \label{tildechi2}
 \tilde{N}\tilde{\chi}^2_{\nu} & \!=\! \left(\tilde{d}_{i}-t_{i}+t_{i}-\bar{t}_i\right)\tilde{\Sigma}^{-1}_{ij}\left(\tilde{d}_{i}-t_{i}+t_{i}-\bar{t}_i\right)\\
& \!=\! \left(\tilde{d}_{i}-t_{i}\right)\tilde{\Sigma}^{-1}_{ij}\left(\tilde{d}_{j}-t_{j}\right)\!+\!\left(t_{i}-\bar{t}_i\right)\tilde{\Sigma}^{-1}_{ij} \left(t_{j}-\bar{t}_{j}\right), \nonumber
\end{align}
where $t_{i}$ is the true value of $t$ computed in $z_{i}$, and in the second line we have omitted cross-product terms whose expectation value is zero because of the independence 
between the data used to fit the model $d_{i}$ and the $\tilde{d}_{i}$ data.%
\footnote{Note that, if $d$ and $\tilde{d}$ are partitions of a  correlated dataset, one is neglecting the 
correlation between these two partitions.} %
The expectation value of the first term is clearly $\tilde{N}$. The expectation value of the second term is 
not trivial:
\begin{align} \label{mean1}
\delta&\!\equiv\!\big\langle\left(t_{i}-\bar{t}_i\right)\tilde{\Sigma}^{-1}_{ij}\left(t_{j}-\bar{t}_{j}\right)\big\rangle \\
&\!=\!\sum_{i}\tilde{\Sigma}^{-1}_{ii}\big\langle\left(t_{i}-\bar{t}_{i}\right)^2\big\rangle+2\sum_{i<j}\tilde{\Sigma}^{-1}_{ij}\big\langle\left(t_{i}-\bar{t}_{i}\right)\left(t_{j}-\bar{t}_{j}\right)\big\rangle\nonumber\,.
\end{align}
In the following we will use the notation that $\bar{t}_{i}$ is a random variable when inside expectation values while 
it is $t(z_i, \{c_{\alpha, {\rm bf}}\})$ when outside, that is, we use the best-fit model in order to estimate the 
true model $t_{i}$.

The first term in the last equation is the variance of $\bar{t}_i$ which can be computed, as in Section~\ref{error}, through a change of variables 
using the covariance matrix on the parameters $\Sigma_{\alpha \beta}$ obtained from equation \eqref{fisher} using the training set:
\begin{align} \label{variancet}
J_{\alpha i} &= \left. \frac{\partial t\left(z_{i},\{c_{\beta}\}\right)}{\partial c_{\alpha}}\right|_{c_{\alpha, {\rm bf}}} %= g_{\alpha}\left(z_i\right) 
= g_{\alpha i}\,, \\
\sigma^2_{t_{i}}&\equiv\big\langle\left(t_{i}-\bar{t}_{i}\right)^2\big\rangle = J_{\alpha i}\Sigma_{\alpha\beta}J_{\beta i}
= g_{\alpha i}\Sigma_{\alpha\beta}g_{\beta i} \,, \nonumber
\end{align}
where we have used equation \eqref{tempe1}.
Note that, thanks to the linearity in the parameters, the best fits $\{c_{\alpha, {\rm bf}}\}$ were not used and this 
computation is exact rather than only valid at the first order in a Taylor expansion.

The second term is,
\begin{align}
\big\langle\left(t_{i}-\bar{t}_{i}\right)\left(t_{j}-\bar{t}_{j}\right)\big\rangle = \big\langle\bar{t}_{i}\bar{t}_{j}\big\rangle - t_{i}t_{j} \,.
\end{align}
Using again equation \eqref{tempe1} one finds:
\begin{align} \label{mean2}
\big\langle\bar{t}_{i}\bar{t}_{j}\big\rangle &= \sum_{\alpha}g_{\alpha i}g_{\alpha j}\big\langle c_{\alpha}^{2}\big\rangle \\
&+ \sum_{\alpha < \beta}\left(g_{\alpha i}g_{\beta j}+g_{\beta i}g_{\alpha j}\right)\big\langle c_{\alpha}c_{\beta}\big\rangle \nonumber\\
&= \sum_{\alpha}g_{\alpha i}g_{\alpha j}\left(\Sigma_{\alpha\alpha}+c_{\alpha, {\rm bf}}^{2}\right)\nonumber\\
&+ \sum_{\alpha < \beta}\left(g_{\alpha i}g_{\beta j}+g_{\beta i}g_{\alpha j}\right)\left(\Sigma_{\alpha\beta}+c_{\alpha, {\rm bf}}\,c_{\beta, {\rm bf}}\right) \nonumber \,,
\end{align}
where we used the best-fit parameters $\{c_{\alpha, {\rm bf}}\}$ in order to estimate the true values of the parameters.
Combining the equations \eqref{mean1}, \eqref{variancet} and \eqref{mean2}, the equation \eqref{tildechi2} can be 
rewritten as:
\begin{align}
\tilde{N}\big\langle\tilde{\chi}^2_{\nu}\big\rangle = \tilde{N} + \delta \,,
\end{align}
where:
\begin{widetext}
\begin{align}
\delta &=\sum_{i}\tilde{\Sigma}^{-1}_{ii}g_{\alpha}\left(z_{i}\right)\Sigma_{\alpha\beta}g_{\beta}\left(z_{i}\right)+2\sum_{i<j}\tilde{\Sigma}^{-1}_{ij}\left[\sum_{\alpha}g_{\alpha i}g_{\alpha j}\left(\Sigma_{\alpha\alpha}+c_{\alpha, {\rm bf}}^{2}\right)+\sum_{\alpha < \beta}\left(g_{\alpha i}g_{\beta j}+g_{\beta i}g_{\alpha j}\right)\left(\Sigma_{\alpha\beta}+c_{\alpha, {\rm bf}}\,c_{\beta, {\rm bf}}\right)-\bar{t}_{i}\bar{t}_{j}\right] . \nonumber
%&=\sum_i  \Sigma^{-1}_{2ii} g_\alpha (z_i) \Sigma_{\alpha \beta} g_\beta (z_i)
%+ 2 \sum_{i<j} \Sigma^{-1}_{2ij}
%\left[ \sum_\alpha g_{\alpha i}g_{\alpha j} \Sigma_{\alpha \alpha}  +\sum_{\alpha < \beta} (g_{\alpha i}g_{\beta j}+g_{\beta i}g_{\alpha j}) \Sigma_{\alpha \beta}   \right ] \nonumber \\
%&=\sum_i  \Sigma^{-1}_{2ii} g_\alpha (z_i) \Sigma_{\alpha \beta} g_\beta (z_i)
%+ 2 \sum_{i<j} \Sigma^{-1}_{2ij}
%\left[ \sum_{\alpha , \beta} (g_{\alpha i}g_{\beta j}) \Sigma_{\alpha \beta}   \right ] \nonumber \\
%&= \sum_{i,j} \Sigma^{-1}_{2ij}
%\left[ \sum_{\alpha , \beta} (g_{\alpha i}g_{\beta j}) \Sigma_{\alpha \beta}   \right ] \nonumber \\
%&= \sum_{i,j,\alpha, \beta} \Sigma^{-1}_{2ij}
%g_{\alpha i}g_{\beta j} \Sigma_{\alpha \beta}  \nonumber \\
%&=  \Sigma_{1\alpha \beta} \Sigma^{-1}_{2\alpha \beta}  .
\end{align}
\end{widetext}
It is then straightforward to obtain that:
\begin{equation}
\delta = \Sigma_{\alpha\beta}\tilde{\Sigma}^{-1}_{\alpha\beta}\,,
\end{equation}
where $\tilde{\Sigma}_{\alpha\beta}$ is the covariance matrix on the parameters obtained from equation \eqref{fisher} using the validation set, that was not used to fit the model.

Motivated by this results, we propose a new generalization of the MSPE:
\begin{equation}
\tilde{\chi}^2_{\delta} = \frac{\left(\tilde{d}_{i}-\bar{t}_{i}\right)\tilde{\Sigma}^{-1}_{ij}\left(\tilde{d}_{j}-\bar{t}_{j}\right)}{\tilde{N}}-\frac{ \Sigma_{\alpha\beta}\tilde{\Sigma}^{-1}_{\alpha\beta}}{\tilde{N}} \,,
\end{equation}
whose expectation value is unity. From the previous equation it follows that for large sets the correction $\delta/\tilde{N}$ should be negligible.
Indeed, as training and validation sets have usually sizes of the same order of magnitude, one has $\delta \approx \alpha_{\rm max}+1$.

\section{\texttt{learning\_curve} package}
\label{sec.lc}

All the learning curves presented in this work were obtained using the package \texttt{learning\_curve}. The package \texttt{learning\_curve} consists of three python scripts to compute and plot learning curves. The three python scripts are the following:
\begin{itemize}
\item \texttt{learning\_curve.py}: general parallelized script for computing learning curves for any linear template function.

\item \texttt{learning\_curve\_linear.py}: script for computing and plotting learning curves for some specific template functions (polynomial, log, inverse or square).

\item \texttt{plot.py}: script for plotting the learning curves from the output files obtained using the script \texttt{learning\_curve.py}.
\end{itemize}

In this work, only the \texttt{learning\_curve\_linear.py} script was used, because its template functions coincide with the template functions adopted in the present analysis. The package is available for download at \href{https://github.com/rodrigovonmarttens/learning\_curve}{github.com/rodrigovonmarttens/learning\_curve}.
%For the detailed instructions about how to use the package see the README.md file.

\end{document}